\documentclass[11pt]{article} 
\usepackage{hyperref}

\usepackage[round]{natbib}
\usepackage{amsbsy}
\usepackage{amsmath}    
\usepackage{amssymb}
\usepackage[figuresright]{rotating}
\usepackage{geometry}
\providecommand{\keywords}[1]{\textbf{\textit{Keywords---}} #1}

\begin{document}

\title{Hypothesis Testing For The Covariance Matrix In High-Dimensional Transposable Data With Kronecker Product Dependence Structure}
\author{
Anestis Touloumis\\ 
Cancer Research UK Cambridge Institute\\
 University of Cambridge\\
 Cambridge CB2 0RE, U.K.\\
\texttt{Anestis.Touloumis@cruk.cam.ac.uk} 
 \and John C. Marioni\\
The EMBL-European Bioinformatics Institute\\
 Hinxton CB10 1SD, U.K.\\
\texttt{marioni@ebi.ac.uk} \and 
Simon Tavar\'e\\ 
Cancer Research UK Cambridge Institute\\
 University of Cambridge\\
 Cambridge CB2 0RE, U.K.\\
\texttt{Simon.Tavare@cruk.cam.ac.uk}
}
\date{}
\maketitle

\begin{abstract}
The matrix-variate normal distribution is a popular model for high-dimensional transposable data because it decomposes the dependence structure of the random matrix into the Kronecker product of two covariance matrices: one for each of the row and column variables. Tests for assessing the sphericity and identity structure of the row (column) covariance matrix in high-dimensional settings while treating the column (row) dependence structure as ``nuisance'' are introduced. The proposed tests are robust to normality departures provided that the Kronecker product dependence structure holds. In simulations, the proposed tests appeared to maintain the nominal level and they tended to be powerful against the alternative hypotheses tested. The utility of the proposed tests is demonstrated by analyzing a microarray study and an electroencephalography study. The proposed testing methodology has been implemented in the R package HDTD. 
\end{abstract}


\section{Introduction}\label{Intro}
Transposable data \citep{Allen2010} refer to matrix-valued random variables where the rows and the columns correspond to two distinct sets of variables of interest. For example, consider the mouse aging atlas project \citep{Zahn2007} where gene expression levels were measured in different tissue samples collected from multiple mice. For each mouse, the data can be organized in a $9 \times 8,932$ matrix where the rows correspond to $9$ different tissues and the columns correspond to $8,932$ genes under study. Here, the two sets of variables are the genes and the different tissues. As a second example, consider a study described by \cite{Zhang1995} in which the electrical activity of the brain was measured using electroencephalography (EEG). Specifically, $64$ electrodes were placed onto the scalp of each subject and the response to visual stimuli was recorded. This procedure was repeated over the course of $256$ time points. A $64 \times 256$ data matrix per subject was then created with row variables corresponding to the electrodes and column variables to the time points. Besides studies in genetics \citep{Allen2010,Allen2012,Efron2009,Teng2009,Yin2012,Ning2011} and EEG studies \cite{Zahn2007,Leng2012}, transposable data arise in spatiotemporal studies \citep{Genton2007,MARDIA1993}, cross-classified multivariate data \citep{Galecki1994,Naik2001}, functional MRI \citep{Allen2010}, financial market targeting \cite{Leng2012} and in time-series \citep{Carvalho2007,Lee2013} among others.

\indent To introduce the notation, consider $N$ independent and identically distributed (i.i.d) transposable $r \times c$ random matrices $\mathbf X_1, \ldots, \mathbf X_N$ such that in each matrix there are $r$ row variables and $c$ column variables. To reflect a high-dimensional setting or equivalently the `small sample size, large number of parameters' paradigm, assume that the sample size $N$ is smaller than or of similar magnitude to the number of observations $r \times c$ in a single matrix. The matrix-variate normal distribution \citep{Dawid1981,Gupta2000} is a popular choice to model transposable data especially in high-dimensional settings \citep{Allen2010,Allen2012,Efron2009,Teng2009,Carvalho2007,Leng2012,Yin2012}. This distribution is defined by three matrix parameters, $\mathbf M$, a matrix of constants, and two positive-definite matrices $\boldsymbol \Sigma_R$ and $\boldsymbol \Sigma_C$. These matrices satisfy the relations $\mathrm{E}[\mathbf X_i]=\mathbf M$ and $\mathrm{cov}[\mathrm{vec}(\mathbf X_i)]=\boldsymbol \Sigma_C \otimes \boldsymbol \Sigma_R$, where $\mathrm{vec}(\mathbf A)$ vectorizes matrix $\mathbf A$ by its columns and $\mathbf A \otimes \mathbf B$ denotes the Kronecker product of the matrices $\mathbf A$ and $\mathbf B$. Therefore, the high-dimensional dependence structure of the transposable data is decomposed into the Kronecker product of two lower-dimensional covariance matrices $\boldsymbol \Sigma_C$ and $\boldsymbol \Sigma_R$, recognized as the covariance matrices of the column and row variables respectively. In the motivating examples, $\boldsymbol \Sigma_R$ describes the dependence structure of different tissues or electrodes and $\boldsymbol \Sigma_C$ the dependence structure of the genes or time-points. 

\indent To the best of our knowledge, no formal procedure exists for performing hypothesis testing for $\boldsymbol \Sigma_R$ in high-dimensional transposable data. Further, employing existing procedures for testing a large covariance matrix based on a sample of random vectors, i.e., $(c=1)$, does not seem appropriate since the potential column-wise dependence structure is ignored. To fill this gap, we consider the problem of hypothesis testing of two structures for $\boldsymbol \Sigma_R$: i) the sphericity structure $\boldsymbol \Sigma_R=\sigma^2 \mathbf I_{r}$, where $\mathbf I_s$ is the identity matrix of size $s$ and $\sigma^2$ is an unknown constant, and ii) the identity structure $\boldsymbol \Sigma_R =\mathbf I_{r}$. To illustrate the practical importance of testing these two hypotheses, suppose that the transposable data are generated from a matrix-variate normal distribution. The sphericity hypothesis for the row covariance matrix imply independence of the row variables in such a way that the transposable data can be written in terms of $r$ independent populations one for each row. In particular, the $a$-th population consists of $N$ $c$-variate random vectors with mean vector the $a$-th row of $\mathbf M$ and covariance matrix $\sigma^{-2} \boldsymbol \Sigma_C$. Therefore, the sphericity hypothesis under the matrix-variate normal model is equivalent to testing simultaneously the hypothesis of independent row variables and of a common covariance matrix structure for the $c$ column variables without making any assumptions about the mean relationship between row and column variables. On the other hand, the primary use of the identity test is to indirectly assess whether a known row covariance matrix $\boldsymbol \Sigma_{R0}$ equals the row-wise covariance structure $\boldsymbol \Sigma_{R}$. To accomplish this, one must apply the transformation $\mathbf X_{i} \longmapsto \boldsymbol \Sigma_{R0}^{-1/2} \mathbf X_{i}$ and then test the identity hypothesis on the transformed random matrices. We also provide two situations in which the identity test can be used directly. First, to provide some statistical evidence regarding the efficiency of the decorrelation algorithm proposed by \cite{Allen2012} in producing independent row and column random variables with unit variance. In this case, the identity test can be used to check whether the transformed row/column variables are indeed independent with unit variance. Second, in studies where transposable data for each subject have been preprocessed in such a way that the measurements across column and/or row variables have sample mean zero and unit variance. Examples of column- and/or doubly-standardized data can be found in microarrays studies \citep{Efron2009}.

\indent The construction of the two proposed test statistics is motivated by the work of \cite{Chen2010a}. In both cases, we estimate a scaled squared Frobenius norm that measures the discrepancy between the null and the alternative hypotheses for $\boldsymbol \Sigma_R$ while treating $\mathbf M$ and $\boldsymbol \Sigma_C$ as `nuisance' matrix parameters. This is reasonable because the squared Frobenius norm of the difference of $\boldsymbol \Sigma_C \otimes \boldsymbol \Sigma_R$ under the sphericity or identity hypothesis and the corresponding alternative hypothesis depends only on the corresponding squared Frobenius norm for $\boldsymbol \Sigma_R$. Next, the unknown parameters of the scaled squared Frobenius norm will be replaced by unbiased and consistent estimators. This allows us to derive the asymptotic distributions of the proposed test statistics and explore their asymptotic power even when the normality assumption does not hold as long as the Kronecker product dependence assumption remains valid. To this end, note that the proposed tests for the row covariance matrix can be applied to the column covariance matrix by interchanging the role of row and column variables. 

\indent This paper is organized as follows. In Section~\ref{kronekermodel}, we describe a nonparametric model for transposable data that preserves the Kronecker product patterned covariance matrix as in the matrix-variate normal distribution. In Section~\ref{TestStatistics}, we specify the working framework that allows us to manage the high-dimensional setting and derive the asymptotic distribution of the proposed test statistic for the identity and sphericity hypotheses of the row (or column) covariance matrix. In Section \ref{Simulation}, we demonstrate the good performance of the proposed tests in simulation studies. In Section~\ref{example}, we apply the test statistics to the motivating datasets. We summarize our findings and discuss future research in Section~\ref{Discussion}. The technical details can be found in the Supplementary Material.

\section{A Nonparametric Model for Transposable Data with Kronecker Dependence Structure} \label{kronekermodel}
Suppose there are $r$ row variables and $c$ column variables and let $\mathbf X_{1},\ldots,\mathbf X_{N}$ be a sample of $N$ i.i.d. $r \times c$ random matrices. As a generative process for the transposable data, assume the non-parametric model
\begin{equation}
\mathbf X_{i} = \boldsymbol \Sigma^{1/2}_R \mathbf Z_{i} \boldsymbol \Sigma^{1/2}_C + \mathbf M,
\label{Nonparametricmodel}
\end{equation}
where $\mathbf M=\mathrm{E}[\mathbf X_i]$ is the $r \times c$ mean matrix, $\boldsymbol \Sigma_m=\boldsymbol \Sigma^{1/2}_m \boldsymbol \Sigma^{1/2}_m$ is a positive definite matrix ($m \in \{R,C\}$), and $\{\mathbf Z_{i}:i=1,\ldots,N\}$ is a sequence of i.i.d. $r \times c$ random matrices. Further, we restrict the moments of the random variables $\{Z_{iab}:a=1,\ldots, r \text{ and }  b=1,\ldots,c\}$ within $\mathbf Z_i$. In particular, we let $\mathrm{E}[Z_{iab}]=0$, $\mathrm{E}[Z^{2}_{iab}]=1$, $\mathrm{E}[Z^4_{iab}]=3+B$ with $-2 \leq B < \infty$, $\mathrm{E}[Z^8_{iab}]<\infty$ and for any positive integers $l_1,\ldots,l_q$ such that $\sum_{\nu=1}^q l_{\nu} \leq 8$ 
\begin{equation}
\mathrm{E}[Z^{l_1}_{ia_1b_1} Z^{l_2}_{ia_2b_2} \ldots Z^{l_q}_{ia_qb_q}]=\mathrm{E}[Z^{l_1}_{ia_1b_1}]\mathrm{E}[Z^{l_2}_{ia_2b_2}] \ldots \mathrm{E}[Z^{l_q}_{ia_qb_q}]
\label{qindependence}
\end{equation}
for $(a_1,b_1)\neq (a_2,b_2) \neq \cdots \neq (a_q,b_q)$. The matrix-variate normal distribution is a special case of model~(\ref{Nonparametricmodel}) obtained if $Z_{iab}$ are i.i.d. $\mathrm{N}(0,1)$ random variables.

\indent The dependence structure of $\mathbf X_{i}$ under model~(\ref{Nonparametricmodel}) satisfies that implied by a matrix-variate normal distribution, that is $\mathrm{cov}[\mathrm{vec}(\mathbf X_{i})]=\boldsymbol \Omega= \boldsymbol \Sigma_C \otimes \boldsymbol \Sigma_R$. Therefore, the dependence structure of the row variables in $\mathbf X_i$ is given by $\boldsymbol \Sigma_R$ and that of the column variables by $\boldsymbol \Sigma_C$. For this reason, we will refer to $\boldsymbol \Sigma_R$ and $\boldsymbol \Sigma_C$ as row and column covariance matrix respectively.
  
\indent The covariance matrices $\boldsymbol \Sigma_R$ and $\boldsymbol \Sigma_C$ are not uniquely identified since $\boldsymbol \Omega=(t\boldsymbol \Sigma_C) \otimes (\boldsymbol \Sigma_R/t)$ for any constant $t>0$. In the context of the matrix-variate normal distribution, this issue has been addressed by either setting a diagonal element of $\boldsymbol \Sigma_C$ equal to 1 \citep{Naik2001,Srivastava2008,Yin2012} or by applying the constraint $\mathrm{tr}(\boldsymbol \Sigma_C)=c$ \citep{MARDIA1993,Theobald2006} where $\mathrm{tr}(\mathbf A)$ denotes the trace of matrix $\mathbf A$. Although neither of these scalings affects the row and column correlation matrices implied by $\boldsymbol \Sigma_R$ and $\boldsymbol \Sigma_C$ respectively, we will adopt the second one because it enables us to construct unbiased and consistent estimators for $\mathrm{tr}(\boldsymbol \Sigma_R)$ and $\mathrm{tr}(\boldsymbol \Sigma^2_R)$ upon which we develop the proposed test statistics. 
\section{Test Statistics}\label{TestStatistics}
Suppose that $\mathbf X_{1},\ldots,\mathbf X_{N}$ are i.i.d. $r \times c$ random matrices satisfying model~(\ref{Nonparametricmodel}) with $\mathrm{tr}(\boldsymbol \Sigma_C)=c$. Consider the sphericity hypothesis test
\begin{equation}
\mathrm{H}_0: \boldsymbol \Sigma_R=\sigma^2 \mathbf I_{r} \text{ vs. } \mathrm{H}_1:\boldsymbol \Sigma_R \neq \sigma^2 \mathbf I_{r},
\label{Sphericity}
\end{equation}
where the unknown constant $\sigma^2>0$ is proportional to the normalizing constant that allows us to identify uniquely the row and column covariance matrices, and the identity hypothesis test
\begin{equation}
\mathrm{H}_0: \boldsymbol \Sigma_R=\mathbf I_{r} \text{ vs. } \mathrm{H}_1:\boldsymbol \Sigma_R \neq \mathbf I_{r}.
\label{Identity}
\end{equation} 

\indent To manage high-dimensional settings, we impose restrictions on the dimension of the row and column covariance matrices. In particular, we assume that as $N \rightarrow \infty$ and $rc=r(N)c(N) \rightarrow \infty$
\begin{equation}
\frac{\mathrm{tr}(\boldsymbol \Sigma_R^4)}{\mathrm{tr}^2(\boldsymbol \Sigma_R^2)}\rightarrow 0  \text{ , } \frac{\mathrm{tr}(\boldsymbol \Sigma_C^4)}{\mathrm{tr}^2(\boldsymbol \Sigma_C^2)}\rightarrow t_1 \geq 0 \text{ and } \frac{\mathrm{tr}(\boldsymbol \Sigma_C^2)}{\mathrm{tr}^2(\boldsymbol \Sigma_C)}\rightarrow t_2 \geq 0,
\label{CovMatAss}
\end{equation}
where $0 \leq t_2 \leq t_1 \leq 1$. Assumption~(\ref{CovMatAss}) does not specify the pairwise limiting ratios of the triplet ($N,r,c$). Hence, it covers both applications in which the sample size might not be expected to increase proportionally to the dimension of the matrix and applications in which $r$ and/or $c$ tend to $\infty$ way faster than $N$ does. This is reflected in the simulation study where the test appeared to behave well in finite sample settings. Instead, assumption~(\ref{CovMatAss}) places mild restrictions on $\boldsymbol \Sigma_R$ and $\boldsymbol \Sigma_C$. Options for the row covariance matrix $\boldsymbol \Sigma_R$ include covariance matrices with eigenvalues bounded away from $0$ and $\infty$, that satisfy a (banded) first-order autoregressive correlation pattern and the variances are bounded \citep{Chen2010a}, or that have a few divergent eigenvalues as long as they diverge slowly \citep{Chen2010}. The restrictions on the column covariance matrix are weaker since $\boldsymbol \Sigma_C$ can also satisfy a compound symmetry correlation pattern provided that the variances of the column variables are bounded away from $0$ and $\infty$. Model~(\ref{Nonparametricmodel}) and assumption (\ref{CovMatAss}) constitute a flexible working framework in order to handle high-dimensional transposable data.

\subsection{Sphericity Test}\label{SphericityStatistics} 
For testing the sphericity hypothesis in~(\ref{Sphericity}), we utilize the scaled squared Frobenius distance
\begin{equation}
\frac{1}{r}\mathrm{tr}\left[\left(\frac{\boldsymbol \Sigma_R}{\mathrm{tr}(\boldsymbol \Sigma_R)/r}-\mathbf I_{r}\right)^2\right]=r\frac{\mathrm{tr}(\boldsymbol \Sigma^2_R)}{\mathrm{tr}^2(\boldsymbol \Sigma_R)}-1.
\label{frobsphericity}
\end{equation}
The parameters $\mathrm{tr}(\boldsymbol \Sigma_R)$ and $\mathrm{tr}(\boldsymbol \Sigma^2_R)$ are estimated by the unbiased and ratio-consistent estimators
\begin{equation*}
T_{1N}=Y_{1N}-Y_{3N}=\frac{1}{c N}\sum_{i=1}^N \mathrm{tr}(\mathbf X_{i}\mathbf X^{\prime}_{i})-\frac{1}{c P^N_2}\sum_{i,j}^{\ast} \mathrm{tr}(\mathbf X_{i}\mathbf X^{\prime}_{j})
\end{equation*}
and 
\begin{align*}
T_{2N}&=Y_{2N}-2Y_{4N}+Y_{5N}\nonumber\\
      &=\frac{1}{c^2 P^N_2}\sum_{i,j}^{\ast} \nolimits \mathrm{tr}(\mathbf X_{i}\mathbf X^{\prime}_{i}\mathbf X_{j}\mathbf X^{\prime}_{j})-2\frac{1}{c^2 P^N_3}\sum_{i,j,k}^{\ast}\nolimits \mathrm{tr}(\mathbf X_{i}\mathbf X^{\prime}_{i}\mathbf X_{j}\mathbf X^{\prime}_{k})\\
      &+\frac{1}{c^2 P^N_4} \sum_{i,j,k,l}^{\ast}\nolimits \mathrm{tr}(\mathbf X_{i}\mathbf X^{\prime}_{j}\mathbf X_{k}\mathbf X^{\prime}_{l})
\end{align*}
respectively, where $P^s_t=s!/(s-t)!$ and $\sum^{\ast}$ denotes summation over mutually distinct indices. Note that the terms in $T_{1N}$ and $T_{2N}$ are $U$-statistics of order two, three and four that are subtracted so that $T_{1N}$ and $T_{2N}$ remain unbiased when $\mathbf M \neq \mathbf 0$. This usage of $U$-statistics was first introduced by \cite{Glasser1961,Glasser1962} and later exploited in the framework of hypothesis testing by \cite{Chen2010a}. Derivations in the Supplementary Material and Theorem~\ref{sphericityvatrhm} show the ratio-consistency property of $T_{1N}$ and $T_{2N}$ under our working framework. 

\indent To construct the test statistic we plug in the unbiased and consistent estimators $T_{1N}$ and $T_{2N}$ in~(\ref{frobsphericity}) to obtain
$$U_N=r\frac{T_{2N}}{T^2_{1N}}-1.$$ Let
\begin{align*}
\sigma^2_U=&\frac{4}{N^2}\left(\frac{\mathrm{tr}(\boldsymbol \Sigma^2_C)}{c^2}\right)^2 +\frac{8}{N} \frac{\mathrm{tr}(\boldsymbol \Sigma^2_C)}{c^2}\mathrm{tr}\left[\left(\frac{\boldsymbol \Sigma^2_R}{\mathrm{tr}(\boldsymbol \Sigma^2_R)}-\frac{\boldsymbol \Sigma_R}{\mathrm{tr}(\boldsymbol \Sigma_R)} \right)^2\right]\nonumber \\
&+\frac{4B}{N} \frac{\mathrm{tr}(\boldsymbol \Sigma_C \circ \boldsymbol \Sigma_C)}{c^2}\mathrm{tr}\left[\left(\frac{\boldsymbol \Sigma_R^2}{\mathrm{tr}(\boldsymbol \Sigma^2_R)}-\frac{\boldsymbol \Sigma_R}{\mathrm{tr}(\boldsymbol \Sigma_R)} \right) \circ \left(\frac{\boldsymbol \Sigma_R^2}{\mathrm{tr}(\boldsymbol \Sigma^2_R)}-\frac{\boldsymbol \Sigma_R}{\mathrm{tr}(\boldsymbol \Sigma_R)} \right) \right]
\end{align*}
where $\mathbf A \circ \mathbf B$ is the Hadamard product of the matrices $\mathbf A$ and $\mathbf B$. Since $-2\leq B$ and  
$\mathrm{tr}(\mathbf A \circ \mathbf A) \leq \mathrm{tr}(\mathbf A^2)$ for any symmetric matrix $\mathbf A$, it follows that $\sigma^2_U>0$. The following theorem provides the limiting distribution of $U_N$ and its proof can be found in the Supplementary Material.
\newtheorem{theorem}{Theorem}
\begin{theorem}
Under model~(\ref{Nonparametricmodel}) and assumption~(\ref{CovMatAss}) 
$$\sigma^{-1}_U \left(\frac{\mathrm{tr}^2(\boldsymbol \Sigma_R)}{\mathrm{tr}(\boldsymbol \Sigma^2_R)}\frac{U_N+1}{r}-1\right)   \stackrel{d}{\rightarrow} N(0,1)$$
where $\stackrel{d}{\rightarrow}$ denotes convergence in distribution.
\label{sphericitythm}
\end{theorem}
Under $\mathrm{H}_0$ in~(\ref{Identity}), $\sigma^2_U$ reduces to $$\sigma^2_{U_0}=\frac{4}{N^2}\left(\frac{\mathrm{tr}(\boldsymbol \Sigma^2_C)}{c^2}\right)^2.$$ In most application, the column-covariance matrix is expected to be unknown and thus the final step to construct the test statistic involves estimation of $\mathrm{tr}(\boldsymbol \Sigma^2_C)$. To do this, we utilize the vectorized form of model~(\ref{Nonparametricmodel}) and write $\mathrm{tr}(\boldsymbol \Sigma^2_C)=\mathrm{tr}(\boldsymbol \Omega^2)/\mathrm{tr}(\boldsymbol \Sigma^2_R)$ where $\boldsymbol \Omega=\boldsymbol \Sigma_C \otimes \boldsymbol \Sigma_R$ is the covariance matrix of $\mathbf R_{i}=\mathrm{vec}(\mathbf X_i)$ for $i=1,\ldots,N$. To estimate $\mathrm{tr}(\boldsymbol \Sigma^2_R)$ we use $T_{2N}$ and to estimate $\mathrm{tr}(\boldsymbol \Omega^2)$ we use 
$$T^{\ast}_{2N}=\frac{1}{P^N_2}\sum_{i,j}^{\ast}\nolimits (\mathbf R^{\prime}_{i}\mathbf R_{j})^2-2\frac{1}{P^N_3}\sum_{i,j,k}^{\ast}\nolimits \mathbf R^{\prime}_{i}\mathbf R_{j}\mathbf R^{\prime}_{i}\mathbf R_{k}+\frac{1}{P^N_4} \sum_{i,j,k,l}^{\ast}\nolimits \mathbf R^{\prime}_{i}\mathbf R_{j}\mathbf R^{\prime}_{k}\mathbf R_{l}.$$ Theorem~\ref{sphericityvatrhm} establishes that $T^{\ast}_{2N}$ is a ratio-consistent estimator of $\mathrm{tr}(\boldsymbol \Omega^2)$ and that $\widehat{\mathrm{tr}(\boldsymbol \Sigma^2_C)}=T^{\ast}_{2N}/T_{2N}$ is a ratio-consistent estimator for $\mathrm{tr}(\boldsymbol \Sigma^2_C)$. The proof of Theorem~\ref{sphericityvatrhm} can be found in the Supplementary Material.
\begin{theorem}
Under model~(\ref{Nonparametricmodel}), assumption~(\ref{CovMatAss}) and either $\mathrm{H}_0$ in~(\ref{Sphericity}) or  $\mathrm{H}_0$ in~(\ref{Identity})
$$\frac{T_{2N}}{\mathrm{tr}(\boldsymbol \Sigma^2_R)} \stackrel{P}{\rightarrow} 1, \frac{T^{\ast}_{2N}}{\mathrm{tr}(\boldsymbol \Omega^2)} \stackrel{P}{\rightarrow} 1 \text{ and } \frac{\widehat{\mathrm{tr}(\boldsymbol \Sigma^2_C)}}{\mathrm{tr}(\boldsymbol \Sigma^2_C)}\stackrel{P}{\rightarrow} 1$$
where $\stackrel{P}{\rightarrow}$ denotes convergence in probability.
\label{sphericityvatrhm}
\end{theorem}
Under $\mathrm{H}_0$ in the sphericity hypothesis~(\ref{Identity}), Theorems~\ref{sphericitythm} and~\ref{sphericityvatrhm} imply that 
$$U^{\ast}_N = \frac{r}{\hat{\sigma}_{U_0}}  U_N \stackrel{d}{\rightarrow} \mathrm{N}(0,1),$$
where $$\widehat{\sigma}_{U_0}=\frac{2}{N}\frac{\widehat{\mathrm{tr}(\boldsymbol \Sigma^2_C)}}{c^2}.$$
Hence, the rejection region of the proposed test at significance level $\alpha$ is $U^{\ast}_N \geq z_{\alpha}$, where $z_{\alpha}$ is the $\alpha$-upper quantile of $\mathrm{N}(0,1)$.

\indent To examine the power function, let $\beta_{U^{\ast}_N}=\Pr(U^{\ast}_N \geq z_{\alpha}| \boldsymbol \Sigma_R \neq \sigma^2 \mathbf I_{r})$. Algebraic manipulation shows that a lower bound of the power function $\beta_{U^{\ast}_N}$ is 
\begin{equation*}
\beta_{U^{\ast}_N} \geq \Phi\left( - \frac{1-\xi_{1N}}{N \xi_{1N}} z_{\alpha}+\sqrt{\frac{c^2}{\mathrm{tr}(\boldsymbol \Sigma^2_C)}}\frac{1}{2\sqrt{\frac{1}{N^2 \xi_{1N}^2}+\psi \frac{\xi_{2N}}{N\xi^2_{1N}}}}\right), 
\end{equation*}
where 
$$\xi_{1N}=1-\frac{\mathrm{tr}^2(\boldsymbol \Sigma_R)}{r \mathrm{tr}(\boldsymbol \Sigma^2_R)} \text{ and } \xi_{2N}=\mathrm{tr}\left[\left(\frac{\boldsymbol \Sigma^2_R}{\mathrm{tr}(\boldsymbol \Sigma^2_R)}-\frac{\boldsymbol \Sigma_R}{\mathrm{tr}(\boldsymbol \Sigma_R)} \right)^2\right]$$
and $\psi=2+\max\{0,B\}$. The consistency of the proposed test is guaranteed as long as $\xi_{2N}/\left({N\xi^2_{1N}}\right) \rightarrow 0$ and $N \xi_{1N}^2 \rightarrow \infty$ or if $\mathrm{tr}(\boldsymbol \Sigma^2_C)/c^2 \rightarrow 0$ and $\xi_{1N}$ and $\xi_{2N}$ converge to some positive constant. The strength of the column-wise correlation might affect the power function of the proposed sphericity test and we explore this further in the simulations. Conditional on the remaining parameters, we expect weak correlation patterns to increase the power of $U^{\ast}_{N}$ since the lower limit of $\beta_{U^{\ast}_N}$ takes its maximum value when $\min\{\mathrm{tr}(\boldsymbol \Sigma^2_C)/c^2\}=1/c=\mathrm{tr}(\mathbf I^2_c)/c^2$.  

\subsection{Identity Test}
In a similar fashion we construct a test statistic for the identity test~(\ref{Identity}). Let
$$V_N=\frac{T_{2N}}{r}-2\frac{T_{1N}}{r}+1$$
be the unbiased estimator of the squared Frobenius norm of $\boldsymbol \Sigma_R - \mathbf I_{r}$ adjusted for the dimension of $\boldsymbol \Sigma_R$, 
\begin{equation*}
\frac{\mathrm{tr}\left[(\boldsymbol \Sigma_R - \mathbf I_{r})^2\right]}{r}=\frac{1}{r}\mathrm{tr}(\boldsymbol \Sigma^2_R) -\frac{2}{r}\mathrm{tr}(\boldsymbol \Sigma_R)+1,
\label{frobidentity}
\end{equation*}
and let 
\begin{align*}
\sigma^2_V =&\frac{4}{N^2}\left(\frac{\mathrm{tr}(\boldsymbol \Sigma^2_C)}{c^2}\right)^2 \mathrm{tr}^2(\boldsymbol \Sigma^2_R)+\frac{8}{N}\frac{\mathrm{tr}(\boldsymbol \Sigma^2_C)}{c^2} \mathrm{tr}\left[(\boldsymbol \Sigma^2_R-\boldsymbol \Sigma_R)^2\right]\\
            &+\frac{4B}{N}\frac{\mathrm{tr}(\boldsymbol \Sigma_C \circ \boldsymbol \Sigma_C)}{c^2} \mathrm{tr}\left[(\boldsymbol \Sigma^2_R-\boldsymbol \Sigma_R) \circ (\boldsymbol \Sigma^2_R-\boldsymbol \Sigma_R)\right]>0.
\end{align*}
The following theorem provides the limiting distribution of $V_N$ and the proof can be found in the Supplementary Material.
\begin{theorem}
Under model~(\ref{Nonparametricmodel}) and assumption~(\ref{CovMatAss})
$$\frac{r V_N-\mathrm{tr}\left[(\boldsymbol \Sigma_R -\mathbf I_{r})^2\right]}{\sigma_V} \stackrel{d}{\rightarrow} \mathrm{N}(0,1).$$
\label{identitythm}
\end{theorem}
Under $\mathrm{H}_0$ in the identity hypothesis~(\ref{Identity}), $\sigma^2_V$ becomes $\sigma^2_{V_0}=\sigma^2_{U_0}r^2$ and consequently, Theorems~\ref{sphericityvatrhm} and~\ref{identitythm} imply that
$$V^{\ast}_N = \frac{r}{\hat{\sigma}_{U_0}}  V_N \stackrel{d}{\rightarrow} \mathrm{N}(0,1).$$
The rejection region of the proposed test at significance level $\alpha$ is $V^{\ast}_N \geq z_{\alpha}$, which implies that $U^{\ast}_N$ and $V^{\ast}_N$ share the same rejection region.

\indent Let $\beta_{V^{\ast}_N}=\Pr(V^{\ast}_N \geq z_{\alpha}| \boldsymbol \Sigma_R \neq \mathbf I_{r})$ be the power function of the proposed test and set $\xi_{3N}=\mathrm{tr}(\boldsymbol \Sigma^2_R)/\left(N\mathrm{tr}\left[(\boldsymbol \Sigma_R-\mathbf I_{r})^2\right]\right)$. Since
\begin{equation*}
\beta_{V^{\ast}_N} \geq \Phi\left( - \frac{r}{\mathrm{tr}(\boldsymbol \Sigma^2_R)} z_{\alpha}+\sqrt{\frac{c^2}{\mathrm{tr}(\boldsymbol \Sigma^2_C)}}\frac{1}{2\sqrt{\xi_{3N}^2+\psi\xi_{3N}}} \right), 
\end{equation*}
it follows that the proposed test is consistent if $\xi_{3N} \rightarrow 0$ or if $\mathrm{tr}(\boldsymbol \Sigma^2_C)/c^2 \rightarrow 0$ and $\xi_{3N}$ converges. 

\indent As mentioned earlier, the identity test can be employed to test the hypothesis $\mathrm{H}_0: \boldsymbol \Sigma_R=\boldsymbol \Sigma_{R0}$ for a known positive definite covariance matrix $\boldsymbol \Sigma_{R0}$ by testing the identity hypothesis~(\ref{Identity}) to the matrices $\boldsymbol \Sigma_{R0}^{-1/2} \mathbf X_{i}$. However, this implies that the trace of the unscaled column covariance matrix, say $\boldsymbol \Sigma^{\ast}_C$, equals its dimension. This will be satisfied the case if $\boldsymbol \Sigma^{\ast}_C$ is a correlation matrix as in \cite{Naik2001,Efron2009}. Otherwise, the following strategy can be adopted for testing $\mathrm{H}_0: \boldsymbol \Sigma_R=\boldsymbol \Sigma_{R0}$. Let 
$$T_{1N^{\ast}}=\frac{1}{N}\sum_{i=1}^N \mathbf R_{i}\mathbf R^{\prime}_{i}-\frac{1}{P^N_2}\sum_{i,j}^{\ast} \mathbf X_{i}\mathbf X^{\prime}_{j}$$
where $\mathbf R_{i}=\mathrm{vec}(\mathbf X_i)$ and note that under $\mathrm{H}_0$ we have that $\mathrm{E}[T_{1N^{\ast}}]=\mathrm{tr}(\boldsymbol \Sigma_{R0})\mathrm{tr}(\boldsymbol \Sigma^{\ast}_C)$. This means that $k=\widehat{\mathrm{tr}(\boldsymbol \Sigma^{\ast}_C)}=T_{1N^{\ast}}/\mathrm{tr}(\boldsymbol \Sigma_{R0})$ is a ratio-consistent estimator of $\mathrm{tr}(\boldsymbol \Sigma^{\ast}_C)$. Therefore, we can test $\mathrm{H}_0: \boldsymbol \Sigma_R=\boldsymbol \Sigma_{R0}$ by testing the identity hypothesis~(\ref{Identity}) to the matrices $\boldsymbol \Sigma_{R0}^{-1/2}\mathbf X_{i}/\sqrt{k} $. 

\subsection{Software availability}
The function \textit{covmat.ts()} of the R/Bioconductor package HDTD implements the proposed sphericity and identity tests. These tests can be applied to either the row or column covariance matrix. The package HDTD is available at \url{http://www.bioconductor.org/packages/3.0/bioc/html/HDTD.html}.  

\section{Remarks}\label{Remarks}
The proposed testing methodology is computationally efficient for three reasons. Firstly, the mean matrix $\mathbf M$ is essentially ignored in the derivation of the test statistics meaning that both $U^{\ast}_N$ and $V^{\ast}_N$ are invariant to the location shift transformation $\mathbf X_i \longmapsto \mathbf X_i-\mathbf M$. Therefore no function of the $r \times c$ mean matrix $\mathbf M$ needs to be estimated and we may assume $\mathbf M=\mathbf 0$ for the rest of the paper, including the proofs in the Supplementary Material. Secondly, no estimation of the `nuisance' covariance matrix parameter $\boldsymbol \Sigma_C$ is required. Instead, we estimate only $\mathrm{tr}(\boldsymbol \Sigma^2_C)$ and thus we avoid estimating the $c(c-1)/2$ non-redundant elements in $\boldsymbol \Sigma_C$, which could be a cumbersome task for large values of $c$. We also confirmed via simulations in Section~\ref{Simulation} that $T^{\ast}_{2N}/T_{2N}$ is a reasonable and accurate estimator of $\mathrm{tr}(\boldsymbol \Sigma^2_C)$. Third, the compuational cost of $U^{\ast}_N$ and $V^{\ast}_N$ can be significantly reduced even if the sample size is large or the dimension of $\boldsymbol \Sigma_R$ is a lot larger than the dimension of $\boldsymbol \Sigma_C$. To accomplish this, one can use the equivalent formulas for $T_{2N}$ and $T^{\ast}_{2N}$ provided in the Supplementary Material and/or the cyclic property of the trace operators when $r > c$. The latter suggests to calculate $T_{1N}$ and $T_{2N}$ based on $c \times c$ matrices, e.g., calculate $T_{1N}$ as $\sum_{i=1}^N \mathrm{tr}(\mathbf X^{\prime}_{i}\mathbf X_{i})/(c N)-\sum_{i,j}^{\ast} \mathrm{tr}(\mathbf X^{\prime}_{j}\mathbf X_{i})/(c P^N_2)$. In the special case of centered transposable data matrices ($\mathbf M=\mathbf 0$), we can calculate the test statistics $V^{\ast}_{N}$ or $U^{\ast}_{N}$ by considering only the first terms in $T_{1N}$, $T_{2N}$ and $T^{\ast}_{2N}$.

\indent An important consequence of model~(\ref{Nonparametricmodel}) is that if we delete any set of row and/or column variables then a Kronecker product dependence structure will still hold for the reduced transposable data. Therefore, the proposed tests can be applied to assess the dependence structure of a smaller set of row variables, a fact that is used repeatedly in analyzing the datasets of the motivating examples in Section~\ref{example}.  

\indent Model~(\ref{Nonparametricmodel}) extends the nonparametric model considered in \cite{Bai1996} and \cite{Chen2010a} to transposable data with a Kronecker product dependence structure. Given this, when a constant $r$-variate mean vector $\boldsymbol \mu$ holds for the row variables ($\mathbf M=\boldsymbol \mu \mathbf 1^{\prime}_{c}$) and the column variables are indeed independent, we expect $U^{*}_{N}$ and $V^{*}_N$ to behave similarly to the corresponding test statistics for the sphericity and the identity hypotheses proposed by \cite{Chen2010a} where these statistics are calculated by treating the $N c$ columns as i.i.d. $r$-variate vectors. However, unlike the tests proposed herein, those in \cite{Chen2010a} do not account for the presence of column-wise dependence structure or of an unrestricted mean matrix $\mathbf M$ even if the column variables are indeed independent.

\indent Finally, if interest lies in applying the sphericity or the identity test to the column covariance matrix, then the transformation $\mathbf X_i \longmapsto \mathbf X^{\prime}_i$ should be performed prior to carrying out the test on the transformed data. In other words, this requires interchanging the role of row and column variables before applying the proposed testing methodology.

\section{Simulations}\label{Simulation}
Simulation studies were performed to investigate the performance of the proposed sphericity test for $\boldsymbol \Sigma_R$. Since the test statistic $U^{\ast}_N$ is invariant to location transformations, we generated i.i.d. random data matrices $\mathbf X_1, \ldots, \mathbf X_N$ assuming that $\mathbf M=\mathbf 0$ in (\ref{Nonparametricmodel}). To examine the nonparametric nature of the test, we simulated under a matrix-variate normal distributional scenario and under a non-normality scenario, in which $\mathbf Z_1, \ldots, \mathbf Z_N$ were simulated such that $Z_{iab}=(Z^{\ast}_{iab}-8)/4$ and $Z^{\ast}_{iab} \stackrel{i.i.d}{\sim} \mathrm{Gamma}(4,0.5)$.

\indent For the triplet $(N,r,c)$ we considered the following settings: $N=20$, $40,$ $60$, $80$, $r=8$, $16$, $32$, $64$, $128$, $256$ and $c=10$, $50$, $100$. These parameters were chosen such that $r \times c$, the number of observations in a single matrix, was no less than the sample size $N$, and thus reflects the high-dimensional settings that motivated the proposed testing procedures.

\indent For the `nuisance' column covariance matrix $\boldsymbol \Sigma_C$, we assumed a first order autoregressive correlation pattern by setting $\boldsymbol \Sigma_C=\{\rho^{|a-b|}\}_{1\leq a,b \leq c}$. To examine the effect of the strength of the column-wise correlation, we used $\rho=0.15$ to reflect a weak correlation pattern and $\rho=0.85$ to reflect a stronger correlation pattern. 

\indent For the row covariance matrix $\boldsymbol \Sigma_R$, we considered the following $4$ configurations:
 \begin{enumerate}
 \item The identity matrix $\boldsymbol \Sigma_R=\mathbf I_{r}$.
 \item A diagonal matrix where the first $r/8$ elements are equal to 2 and the remainder are equal to $1$. This dependence structure implies heteroskedastic row variables.
 \item A compound symmetry covariance matrix in which $\boldsymbol \Sigma_R=0.9 \mathbf I_{r} + 0.2 \mathbf 1_{r} \mathbf 1^{\prime}_{r}$.
 \item A tridiagonal correlation matrix in which the non-zero off-diagonal elements are equal to $0.1$.
 \end{enumerate}
In each simulation scheme, we used 1000 replicates and we calculated the proportions of rejections based on $U^{\ast}_N$ at a $5\%$ nominal significance level. The empirical level of the proposed test was calculated when $\boldsymbol \Sigma_R=\mathbf I_{r}$ while the other three configurations of $\boldsymbol \Sigma_R$ were used to estimate the empirical power. 

\indent Simulation results and a more descriptive version of the simulation findings can be found in the Supplementary Material. In summary, we noticed that the empirical level of the sphericity test well approximated the nominal level, especially when the number of row and column variables $(rc)$ increased. We noticed that the sample size can become an important factor for maintaining the nominal size only when the values of $r$ and/or $c$ are small and the column variables are weakly correlated. In terms of the power, we noticed that the test was extremely powerful unless $\boldsymbol \Sigma_R$ satisfied the tridiagonal correlation matrix, $\rho=0.85$ and $c = 10$. As expected, the power of the test approached one as soon as $c=50$ and $N=40$. We did not observe a distributional effect in the empirical size or power, which verifies the nonparametric nature of the test. In addition, we explored the performance of the proposed identity test in the above settings. As expected, the identity test rejects more often the null hypothesis than the sphericity test but any discrepancy in the size and power diminished as $N \rightarrow \infty$ and $rc \rightarrow \infty$. Finally, it appears that $\widehat{\mathrm{tr}(\boldsymbol \Sigma_C^2)}$ is an accurate estimator of the `nuisance' parameter $\mathrm{tr}(\boldsymbol \Sigma^2_C)$, as desired.

\section{Numerical Examples}\label{example}
\subsection{Mouse aging project dataset}
In a project to study aging in mice, \cite{Zahn2007} measured gene expression levels for 8,932 genes in up to 16 tissues per mouse ($N=40$). Herein we focus on the subset of genes ($c=46$) that play a role in the mouse endothelial growth factor (VEGF) signaling pathway, and investigate their expression levels across $r=9$ tissues, namely the adrenal glands, cerebrum, hippocampus, kidney, lung, muscle, spinal cord, spleen and thymus. \cite{Yin2012} and \cite{Ning2011} have previously analyzed the VEGF signaling pathway using slightly different subsets of the original dataset. As pointed out by \cite{Ning2011}, the Kronecker product form for the dependence structure is plausible, but the quantile-quantile plots do not seem to support a normality assumption for the transposable data.

\indent An important aspect of the VEGF signaling pathway dataset is to infer the dependence structure among the $9$ tissues. To do this, we adopted a simple approach that allows us to identify pairwise tissue correlations that might be statistically significant. First, we estimated the tissue covariance matrix using the sample analogue
$$\widehat{\boldsymbol \Sigma}_R=\frac{1}{(N-1)c}\sum_{i=1}^{N}(\mathbf X_i-\bar{\mathbf X})^{\prime}(\mathbf X_i-\bar{\mathbf X})$$
where $\bar{\mathbf X}=\sum_{i=1}^N \mathbf X_i/N$ is the sample mean matrix. Note that $\widehat{\boldsymbol \Sigma}_R$ is an unbiased estimator of the row covariance matrix under the constraint $\mathrm{tr}(\boldsymbol \Sigma_C)$. Based on the correlation matrix implied by $\widehat{\boldsymbol \Sigma}_R$, we found that the estimated pairwise correlation parameters ranged from $-0.139$ to $0.374$. This indicates a weak to moderate correlation pattern among the tissues. In fact, there are only four tissue pairs with estimated correlation parameters larger than $0.1$ in absolute value: (i) lung-spinal cord, (ii) hippocampus-kidney, (iii) cerebrum-spleen and (iv) cerebrum-thymus. Unlike the tissue graphical network in \cite{Ning2011}, \cite{Yin2012} concluded that these tissue pairs are connected. If we ignore the tissues of the first two pairs, then we fail to reject the sphericity hypothesis for the truncated tissue covariance matrix at a $5\%$ significance level ($U^{\ast}_N=1.281$ and $p$-value$=0.1$). At the same time, we reject the sphericity hypothesis for the tissue covariance matrix when only the lung, spinal cord, hippocampus and kidney tissues are considered ($U^{\ast}_N=17.6411$ and $p$-value$<0.001$). The same inferential analysis holds after applying a Bonferonni correction for multiple testing. These results imply that the regulation of the VEGF signaling pathway is uncorrelated across the adrenal glands, cerebrum, muscle, spleen and thymus, and thus suggesting that previous work might have overestimated the strength of these tissue dependencies in mice \citep{Yin2012,Ning2011}. 

\subsection{EEG dataset}
The EEG dataset \citep{Zhang1995}, available at \url{http://kdd.ics.uci.edu/databases/eeg/eeg.data.html}, describes a study that explores whether EEG correlates alcoholism with genetic predisposition. The $122$ subjects who participated in this study were classified into either an alcoholic or a control group. For each subject, voltage fluctuations were recorded from $64$ electrodes placed on the subject's scalp. Each subject was shown either one stimulus or two (matched or unmatched) stimuli and the voltage measures were recorded at $256$ consecutive time points. This procedure was then repeated for up to $120$ trials. We consider data from the alcoholic group and for each of the $77$ subjects, we created a two-dimensional data matrix such that the rows correspond to the $64$ electrodes, the columns to the $256$ time points and the values represent the average of the corresponding voltage measures across the available number of trials.
  
\indent Our goal was to assess the likelihood that the electrodes were uncorrelated and sharing the same covariance and/or mean structure. For testing the mean structure, we applied the testing procedure proposed by \cite{Touloumis2014} while to assess the assumption of a common covariance matrix for the electrodes, we used the proposed sphericity and identity tests. The sphericity and the identity hypothesis for the covariance matrix of the electrodes were rejected ($U_N^{\ast}=1793.833$ and $V_N^{\ast}=82178.41$) and so did the hypothesis of a common mean vector for the electrodes. As a follow-up study, we used the spatial information that $58$ out of $64$ electrodes belong to five specific regions of the brain (central, parietal, occipital, frontal and temporal). In each of these regions, we rejected the sphericity and identity hypotheses for the dependence structure and the hypothesis of a common mean structure. These imply that it is not appropriate to treat the electrodes as independent or uncorrelated random variables with the same covariance and/or mean structure even if we restrict our attention to a specific region of the brain. Therefore, statistical analysis of this dataset should consider the structural information of the transposable data and requires careful modeling of the mean structure and of the dependence among the electrodes.

\section{Discussion}\label{Discussion}
We considered novel test statistics for assessing the sphericity and the identity hypothesis for a row (or column) covariance matrix in high-dimensional transposable data, conditional upon the $N$ i.i.d. random matrices having a Kronecker product dependence structure. Our test statistics are robust to departures from the popular matrix-variate normal model and computationally inexpensive as shown in the Supplementary Material. From a theoretical perspective, the high-dimensional setting is handled by restricting the form of the row and column covariance matrices under consideration. This class of covariance matrices includes many dependence structures of interest with more flexibility possible for the column (i.e., `nuisance') covariance matrix. The proposed tests appeared to maintain the nominal size while being powerful against the alternatives tested. The proposed methodology is implemented in the R package HDTD.  

\indent The appropriateness of the Kronecker product dependence structure in transposable data should be explored before applying the proposed test statistics. Relevant literature \citep{Dutilleul1999,Mitchell2005,Mitchell2006,Roy2005} is limited to likelihood ratio test statistics under a normality assumption for the vectorized form of the transposable data when $(rc)<N$. Since these tests cannot be used in high-dimensional settings, \cite{Yin2012} and \cite{Ning2011} proposed empirical approaches to examine the validity of the Kronecker product dependence structure. In particular, the strategy of \cite{Ning2011} can be applied in the setting we consider. Under model~(\ref{Nonparametricmodel}), the covariance between two random variables $X_{ia_1b_1}$ and $X_{ia_2b_2}$ in $\mathbf X_i$ is $\mathrm{Cov}[X_{ia_1b_1},X_{ia_2b_2}]=\Sigma^R_{a_1a_2} \Sigma^C_{b_1b_2}$ where $\Sigma^R_{ab}$ is the $(a,b)$th element of $\boldsymbol \Sigma_R$ and $\Sigma^C_{ab}$ is the $(a,b)$th element of $\boldsymbol \Sigma_C$. Consequently the correlation of $X_{ia_1b_1}$ and $X_{ia_2b_2}$ is
\begin{equation} \label{CorrStructure}
\mathrm{Corr}[X_{ia_1b_1},X_{ia_2b_2}] = \mathrm{Corr}[X_{ia_1s},X_{ia_2s}] \mathrm{Corr}[X_{ikb_1},X_{ikb_2}], 
\end{equation}
for $k=1,\ldots,r$ and $s=1,\ldots,c$. If we vectorize the $N$ random matrices and estimate all possible correlations, then we would expect relation~(\ref{CorrStructure}) to hold when we plug-in the corresponding estimators of the correlations. However, this procedure might be computationally intensive and sensitive to high-dimensional settings when it comes to estimating the correlation parameters. In future research, we plan to develop a rigorous testing procedure to assess the Kronecker product structure in high-dimensional transposable data under the nonparametric model~(\ref{Nonparametricmodel}).

\bibliographystyle{plainnat}
\bibliography{dissrefer.bib}

\end{document}

\title{Supplementary Materials: Hypothesis Testing For The Covariance Matrix In High-Dimensional Transposable Data With Kronecker Product Dependence Structure}
\author{
Anestis Touloumis\\ 
Cancer Research UK Cambridge Institute\\
 University of Cambridge\\
 Cambridge CB2 0RE, U.K.\\
\texttt{Anestis.Touloumis@cruk.cam.ac.uk} 
 \and John C. Marioni\\
The EMBL-European Bioinformatics Institute\\
 Hinxton CB10 1SD, U.K.\\
\texttt{marioni@ebi.ac.uk} \and 
Simon Tavar\'e\\ 
Cancer Research UK Cambridge Institute\\
 University of Cambridge\\
 Cambridge CB2 0RE, U.K.\\
\texttt{Simon.Tavare@cruk.cam.ac.uk}
}
\date{}
\maketitle

%\thankstext{t1}{Some comment}
%\thankstext{t2}{First supporter of the project}
%\thankstext{t3}{Second supporter of the project}

\section{Alternative formulas}
Algebraic manipulation shows that 
\begin{align*}
T_{2N}  &=Y_{2N}-2Y_{4N}+Y_{5N}\nonumber\\ 
        &=\frac{1}{c^2 P^N_2}\sum_{i,j}^{\ast} \nolimits \mathrm{tr}(\mathbf X_{i}\mathbf X^{\prime}_{i}\mathbf X_{j}\mathbf X^{\prime}_{j})-2\frac{1}{c^2 P^N_3}\sum_{i,j,k}^{\ast}\nolimits \mathrm{tr}(\mathbf X_{i}\mathbf X^{\prime}_{i}\mathbf X_{j}\mathbf X^{\prime}_{k})\\
        &+\frac{1}{c^2 P^N_4} \sum_{i,j,k,l}^{\ast}\nolimits \mathrm{tr}(\mathbf X_{i}\mathbf X^{\prime}_{j}\mathbf X_{k}\mathbf X^{\prime}_{l})\\
        &=\frac{1}{c^2 P^N_2}Y^{\star}_{2N}-2\frac{1}{c^2 P^N_3}Y^{\star}_{4N}+\frac{1}{c^2 P^N_4}Y^{\star}_{5N}\nonumber\\     
\end{align*}
where 
\begin{align*}
Y^{\star}_{2N}  &=\sum_{i,j}^{\ast}\mathrm{tr}(\mathbf X_{i}\mathbf X^{\prime}_{i}\mathbf X_{j}\mathbf X^{\prime}_{j})\\ 
\end{align*}
and 
\begin{align*}
Y^{\star}_{4N}  &= N^2 Y^{\star}_{41N}-(N-1)^2 Y^{\star}_{42N}-Y^{\star}_{2N}+2(N-1)Y^{\star}_{43N}\\
Y^{\star}_{41N} &=\sum_{i}\mathrm{tr}\left[(\mathbf X_{i}-\bar{\mathbf X})(\mathbf X_{i}-\bar{\mathbf X})^{\prime}\mathbf X_{i} \mathbf X_{i}^{\prime}\right]\\
\bar{\mathbf X} &=\sum_{i}\mathbf X_{i}/N\\
Y^{\star}_{42N} &=\sum_{i}\mathrm{tr}(\mathbf X_{i} \mathbf X_{i}^{\prime}\mathbf X_{i} \mathbf X_{i}^{\prime})\nonumber\\
Y^{\star}_{43N} &=\sum_{i,j}^{\ast}\mathrm{tr}(\mathbf X_{i} \mathbf X_{i}^{\prime}\mathbf X_{i} \mathbf X_{j}^{\prime})\\
\end{align*}
and
\begin{align*}
Y^{\star}_{5N}  &= \frac{1}{3}\left[(N-1)(N^2-3N+3)Y^{\star}_{42N}+(2N-3)(Y^{\star}_{2N}+Y^{\star}_{52N}+Y^{\star}_{53N}) \right. \\
                &\left. {} + 2(N-3)(Y^{\star}_{4N}+Y^{\star}_{54N}+Y^{\star}_{55N})-4(N^2-3N+3)Y^{\star}_{43N}-N^2Y^{\star}_{51N} \right]\\
Y^{\star}_{51N} &=\sum_{i}\mathrm{tr}\left[(\mathbf X_{i}-\bar{\mathbf X})(\mathbf X_{i}-\bar{\mathbf X})^{\prime}(\mathbf X_{i}-\bar{\mathbf X})(\mathbf X_{i}-\bar{\mathbf X})^{\prime}\right]\\
Y^{\star}_{52N} &=\sum_{i,j}^{\ast}\mathrm{tr}(\mathbf X_{i} \mathbf X_{j}^{\prime}\mathbf X_{j} \mathbf X_{i}^{\prime})\\
Y^{\star}_{53N} &=\sum_{i,j}^{\ast}\mathrm{tr}(\mathbf X_{i} \mathbf X_{j}^{\prime}\mathbf X_{i} \mathbf X_{j}^{\prime})\\
Y^{\star}_{54N} &=\sum_{i,j,k}^{\ast}\nolimits \mathrm{tr}(\mathbf X_{i}\mathbf X^{\prime}_{j}\mathbf X_{k}\mathbf X^{\prime}_{i})\\
                &=N^2Y^{\star}_{541N}+2(N-1)Y^{\star}_{43N}-(N-1)^2Y^{\star}_{42N}-Y^{\star}_{52N}\\
Y^{\star}_{541N} &=\sum_{i}\mathrm{tr}\left[(\mathbf X_{i}-\bar{\mathbf X})\mathbf X^{\prime}_{i} \mathbf X_{i}(\mathbf X_{i}-\bar{\mathbf X})^{\prime}\right]\\
Y^{\star}_{55N} &=\sum_{i,j,k}^{\ast}\nolimits \mathrm{tr}(\mathbf X_{i}\mathbf X^{\prime}_{j}\mathbf X_{i}\mathbf X^{\prime}_{k})\\
                &=N^2Y^{\star}_{551N}+2(N-1)Y^{\star}_{43N}-(N-1)^2Y^{\star}_{42N}-Y^{\star}_{53N}\\
Y^{\star}_{551N} &=\sum_{i}\mathrm{tr}\left[\mathbf X_{i}(\mathbf X_{i}-\bar{\mathbf X})^{\prime} \mathbf X_{i}(\mathbf X_{i}-\bar{\mathbf X})^{\prime}\right]\\
\end{align*}
Note that the  cyclic property should be used if $r>c$. Using the results from \cite{Himenoa2012}, it follows that
\begin{align*}
T^{\ast}_{2N} &=\frac{1}{P^N_2}\sum_{i,j}^{\ast} (\mathbf R^{T}_{i}\mathbf R_{j})^2-2\frac{1}{P^N_3}\sum_{i,j,k}^{\ast} \mathbf R^{T}_{i}\mathbf R_{j}\mathbf R^{T}_{i}\mathbf R_{k}+\frac{1}{P^N_4} \sum_{i,j,k,l}^{\ast} \mathbf R_{i}\mathbf R^{T}_{j}\mathbf R_{k}\mathbf R^{T}_{l}\\
			&= \frac{N-1}{N(N-2)(N-3)}\left[(N-1)(N-2)\mathrm{tr}(\mathbf S^2)+\mathrm{tr}^2(\mathbf S)-NQ\right]
\end{align*}
where $$Q=\frac{1}{N-1}\sum_{i=1}^N \left[(\mathbf{R}_i-\bar{\mathbf{R}})^T(\mathbf{R}_i-\bar{\mathbf{R}})\right]^2$$
and $\bar{\mathbf R}=\sum_{i}\mathbf R_{i}/N$. The equivalent forms of $T_{2N}$ and $T^{\ast}_{2N}$ imply that the computational cost of the proposed statistics reduces from $O(N^4)$ to $O(N^2)$.

\section{Useful Identities}
We list four properties of the Kronecker and Hadamard product (P1-P4) and five results (P5-P9) under the nonparametric model (2.1) with $\mathbf M=\mathbf 0$ because the test statistics $V^{\ast}_{N}$ and $U^{\ast}_{N}$ are invariant to location transformations.
\begin{itemize}
\item [P1:]$\mathrm{\mathrm{\mathrm{tr}}}(\mathbf {A^{\prime}BCD^{\prime}})=\mathrm{vec}(\mathbf A)^{\prime}(\mathbf D \otimes \mathbf B) \mathrm{vec}(\mathbf C)$.
\item [P2:]$\mathrm{tr}(\mathbf A^p \otimes \mathbf B^q)= \mathrm{tr}(\mathbf A^p) \mathrm{tr}(\mathbf B^q)$ for $p,q=1,2,3\ldots$
\item [P3:]$\mathrm{tr}\left[(\mathbf A \otimes \mathbf B) \circ (\mathbf A \otimes \mathbf B)\right]= \mathrm{tr}(\mathbf A \circ \mathbf A) \mathrm{tr}(\mathbf B \circ \mathbf B) $.
\item [P4:]$\mathrm{vec}(\mathbf A \mathbf B\mathbf C^{\prime})=(\mathbf{C} \otimes \mathbf{A}) \mathrm{vec}(\mathbf{B})$.
\item [P5:]$\mathrm{E}[\mathbf Z_{i} \mathbf B_2 \mathbf Z^{\prime}_{i}]= \mathrm{tr}(\mathbf B_2) \mathbf I_{r}$.
\item [P6:]$\mathrm{E}[\mathbf Z^{\prime}_{i} \mathbf B_1 \mathbf Z_{i}]= \mathrm{tr}(\mathbf B_1) \mathbf I_{c}$.
\item [P7:]$\mathrm{E}[\mathrm{tr}^2(\mathbf B_1 \mathbf Z_{i} \mathbf B_2 \mathbf Z^{\prime}_{j})]= \mathrm{tr}(\mathbf B^2_1) \mathrm{tr}(\mathbf B^2_2)$.
\item [P8:]$\mathrm{E}[\mathrm{tr}(\mathbf Z^{\prime}_{i} \mathbf B_1 \mathbf Z_{i}  \mathbf B_2 \mathbf Z^{\prime}_{i} \mathbf B_1 \mathbf Z_{i} \mathbf B_3)]= \mathrm{tr}^2(\mathbf B_1)\mathrm{tr}(\mathbf B_2\mathbf B_3)+\mathrm{tr}(\mathbf B^2_1)\mathrm{tr}(\mathbf B_2\mathbf B_3)+\mathrm{tr}(\mathbf B^2_1)\mathrm{tr}(\mathbf B_2)\mathrm{tr}(\mathbf B_3)+B \mathrm{tr}(\mathbf B_1 \circ \mathbf B_1)\mathrm{tr}(\mathbf B_2 \circ \mathbf B_3)$.
\item [P9:]$\mathrm{E}\left[\mathrm{tr}\left[(\mathbf B_1 \mathbf Z_{i} \mathbf B_2 \mathbf Z^\prime_{i} \mathbf B_1) \circ (\mathbf B_1 \mathbf Z_{i} \mathbf B_2 \mathbf Z^\prime_{i} \mathbf B_1) \right]\right]= B \mathrm{tr}(\mathbf B_2 \circ \mathbf B_2) \sum_{a,b} B^4_{1,ab}+ [2\mathrm{tr}(\mathbf B_2^2)+\mathrm{tr}^2(\mathbf B_2)] \mathrm{tr}(\mathbf B_1^2 \circ \mathbf B_1^2)$, where $B^4_{1,ab}$ is the $(a,b)$-th element of $\mathbf B^4_1$.
\end{itemize}
In the above, it is assumed that the dimensions of the involved matrices are meaningful for each of the operations considered, the matrices $\mathbf B_1$, $\mathbf B_2$ and $\mathbf B_3$ are symmetric and that the elements of $\mathbf Z_i$ satisfy the moment restrictions defined bellow model (2.1). 

\section{Moment derivations}\label{Moment Derivations}
We derive the first two moments for the $U$-statistics in $T_{1N}$ and $T_{2N}$. First note that $\mathrm{E}[Y_{1N}]=\mathrm{tr}(\boldsymbol \Sigma_R)$, $\mathrm{E}[Y_{2N}]=\mathrm{tr}(\boldsymbol \Sigma^2_R)$ and $\mathrm{E}[Y_{3N}]=\mathrm{E}[Y_{4N}]=\mathrm{E}[Y_{5N}]=0$. Now
\begin{align*}
\mathrm{E}[Y^{2}_{1N}]&=\frac{1}{c^2 N} \left\{\mathrm{E}[\mathrm{tr}^2(\mathbf X_i\mathbf X^{\prime}_i)]+(N-1)\mathrm{E}[\mathrm{tr}(\mathbf X_i\mathbf X^{\prime}_i)\mathrm{tr}(\mathbf X_j\mathbf X^{\prime}_j)] \right\}\\
             &=\mathrm{tr}^2(\boldsymbol \Sigma_R)+\frac{2}{N}\frac{\mathrm{tr}(\boldsymbol \Sigma^2_C)}{c^2}\mathrm{tr}(\boldsymbol \Sigma^2_R)+\frac{B}{N}\frac{\mathrm{tr}(\boldsymbol \Sigma_C \circ \boldsymbol \Sigma_C)}{c^2} \mathrm{tr}(\boldsymbol \Sigma_R \circ \boldsymbol \Sigma_R),\\
\mathrm{E}[Y^{2}_{3N}]&=\frac{2}{c^2 N(N-1)}\mathrm{E}[\mathrm{tr}^2(\mathbf X_i\mathbf X^{\prime}_j)]=\frac{2}{N(N-1)}\frac{\mathrm{tr}(\boldsymbol \Sigma^2_C)}{c^2}\mathrm{tr}(\boldsymbol \Sigma^2_R),
\end{align*}
\begin{align*}
\mathrm{E}[Y^2_{2N}] =&\frac{2}{c^4 P^N_2} \mathrm{E}[\mathrm{tr}^2(\mathbf X_i\mathbf X^{\prime}_i\mathbf X_j\mathbf X^{\prime}_j)]\\
                      &+\frac{(N-2)(N-3)}{c^4 P^N_2}\mathrm{E}[\mathrm{tr}(\mathbf X_i\mathbf X^{\prime}_i\mathbf X_j\mathbf X^{\prime}_j)\mathrm{tr}(\mathbf X_k\mathbf X^{\prime}_k\mathbf X_l\mathbf X^{\prime}_l)]\\
                      &+\frac{4(N-2)}{c^4 P^N_2}\mathrm{E}[\mathrm{tr}(\mathbf X_i\mathbf X^{\prime}_i\mathbf X_j\mathbf X^{\prime}_j)\mathrm{tr}(\mathbf X_i\mathbf X^{\prime}_i\mathbf X_k\mathbf X^{\prime}_k)] \\
%            &=\frac{1}{c^4 P^N_2}\{2\mathrm{E}[\mathrm{tr}^2(\mathbf X_i\mathbf X^{\prime}_i\mathbf X_j\mathbf X^{\prime}_j)]+(N-2)(N-3) c^4 \mathrm{tr}^2(\boldsymbol \Sigma^2_R)\\
%            &+4(N-2) \left[ c^4 \mathrm{tr}^2(\boldsymbol \Sigma^2_R)+ c^2 \mathrm{tr}(\boldsymbol \Sigma^2_C) \mathrm{tr}(\boldsymbol \Sigma^4_R)+B c^2 \mathrm{tr}(\boldsymbol \Sigma_C \circ \boldsymbol \Sigma_C)\mathrm{tr}(\boldsymbol \Sigma^2_R \circ \boldsymbol \Sigma^2_R)\right]\}\\
%            &=\frac{1}{c^4 P^N_2}\{2\mathrm{tr}^2(\boldsymbol \Sigma^2_R)\mathrm{E}[\mathrm{tr}^2(\mathbf Z_{j} \boldsymbol \Sigma_C \mathbf Z_{j}^{\prime} \boldsymbol \Sigma_R^2)]+4 \mathrm{tr}^2(\boldsymbol \Sigma_C^2)\mathrm{E}[\mathrm{tr}(\mathbf Z^{\prime}_{i} \boldsymbol \Sigma_C \mathbf Z_{i}  \boldsymbol \Sigma^2_R \mathbf Z^{\prime}_{i} \boldsymbol \Sigma_C \mathbf Z_{i} \boldsymbol \Sigma_R^2)]\\
%            &+2 B \mathrm{tr}(\boldsymbol \Sigma_C \circ \boldsymbol \Sigma_C)\mathrm{E}[\mathrm{tr}(\boldsymbol \Sigma_R \mathbf Z_i \boldsymbol \Sigma_C \mathbf Z^{\prime}_i \boldsymbol \Sigma_R \circ \boldsymbol \Sigma_R \mathbf Z_i \boldsymbol \Sigma_C \mathbf Z^{\prime}_i \boldsymbol \Sigma_R)]\\
%            &+(N-2)(N+1) c^4 \mathrm{tr}^2(\boldsymbol \Sigma^2_R)+4(N-2) \left[c^2 \mathrm{tr}(\boldsymbol \Sigma^2_C) \mathrm{tr}(\boldsymbol \Sigma^4_R)+B c^2 \mathrm{tr}(\boldsymbol \Sigma_C \circ \boldsymbol \Sigma_C)\mathrm{tr}(\boldsymbol \Sigma^2_R \circ \boldsymbol \Sigma^2_R)\right]\}\\
            =&\mathrm{tr}^2(\boldsymbol \Sigma_R^2)+\frac{8}{N}\frac{\mathrm{tr}(\boldsymbol \Sigma_C^2)}{c^2}\mathrm{tr}(\boldsymbol \Sigma_R^4)+\frac{4}{P^N_2}\frac{\mathrm{tr}^2(\boldsymbol \Sigma_C^2)}{c^4} \left[\mathrm{tr}^2(\boldsymbol \Sigma_R^2)+\mathrm{tr}(\boldsymbol \Sigma_R^4)\right]\\
            &+\frac{4B}{N}\frac{\mathrm{tr}(\boldsymbol \Sigma_C \circ \boldsymbol \Sigma_C)}{c^2}\mathrm{tr}(\boldsymbol \Sigma^2_R \circ \boldsymbol \Sigma^2_R)\\
            &+\frac{4B}{P^N_2}\frac{\mathrm{tr}(\boldsymbol \Sigma_C \circ \boldsymbol \Sigma_C)}{c^2}\frac{\mathrm{tr}(\boldsymbol \Sigma^2_C)}{c^2}\mathrm{tr}(\boldsymbol \Sigma^2_R \circ \boldsymbol \Sigma^2_R)\\
            &+\frac{2B^2}{N(N-1)}\frac{\mathrm{tr}^2(\boldsymbol \Sigma_C \circ \boldsymbol \Sigma_C)}{c^4}\sum_{a,b} \Sigma^4_{R,ab},
\end{align*}
\begin{align*}
\mathrm{E}[Y^2_{4N}] =&\frac{2}{c^4 P^N_3}\mathrm{E}[\mathrm{tr}^2(\mathbf X_i\mathbf X^{\prime}_i\mathbf X_j\mathbf X^{\prime}_k)]\\
&+\frac{2(N-3)}{c^4 P^N_3}\mathrm{E}[\mathrm{tr}(\mathbf X_i\mathbf X^{\prime}_i\mathbf X_j\mathbf X^{\prime}_k)\mathrm{tr}(\mathbf X_l\mathbf X^{\prime}_l\mathbf X_j\mathbf X^{\prime}_k)]\\
%            &=\frac{1}{c^4 P^N_3}\{2\left(c^2\mathrm{tr}(\boldsymbol \Sigma^2_C)\mathrm{tr}(\boldsymbol \Sigma^4_R)+\mathrm{tr}^2(\boldsymbol \Sigma^2_C)(\mathrm{tr}(\boldsymbol \Sigma^4_R)+\mathrm{tr}^2(\boldsymbol \Sigma^2_R))+B \mathrm{tr}(\boldsymbol \Sigma_C \circ \boldsymbol \Sigma_C) \mathrm{tr}(\boldsymbol \Sigma^2_R \circ \boldsymbol \Sigma^2_R)\right)\}\\
%            &=\frac{1}{c^4 P^N_2}\{2\mathrm{E}[\mathrm{tr}^2(\mathbf X_i\mathbf X^{\prime}_i\mathbf X_j\mathbf X^{\prime}_j)]+(N-2)(N-3) c^4 \mathrm{tr}^2(\boldsymbol \Sigma^2_R)\\
%            &+4(N-2) \left[c^4 \mathrm{tr}^2(\boldsymbol \Sigma^2_R)+ c^2 \mathrm{tr}(\boldsymbol \Sigma^2_C) \mathrm{tr}(\boldsymbol \Sigma^4_R)+B c^2 \mathrm{tr}(\boldsymbol \Sigma_C \circ \boldsymbol \Sigma_C)\mathrm{tr}(\boldsymbol \Sigma^2_R \circ \boldsymbol \Sigma^2_R)\right]\\
%            &+2(N-3)c^2 \mathrm{tr}(\boldsymbol \Sigma^2_C) \mathrm{tr}(\boldsymbol \Sigma^4_R)\}\\
            =&\frac{2}{N(N-1)}\frac{\mathrm{tr}(\boldsymbol \Sigma^2_C)}{c^2}\left\{\mathrm{tr}(\boldsymbol \Sigma^4_R)+\frac{\mathrm{tr}(\boldsymbol \Sigma^2_C)}{c^2}\frac{\mathrm{tr}^2(\boldsymbol \Sigma^2_R)+\mathrm{tr}(\boldsymbol \Sigma^4_R)}{(N-2)(N-3)}\right\}\\
             &+\frac{2B}{P^N_3}\frac{\mathrm{tr}(\boldsymbol \Sigma^2_C)}{c^2}\frac{\mathrm{tr}(\boldsymbol \Sigma_C \circ \boldsymbol \Sigma_C)}{c^2}\mathrm{tr}(\boldsymbol \Sigma^2_R \circ \boldsymbol \Sigma^2_R),
\end{align*}
and
\begin{align*}
\mathrm{E}[Y^2_{5N}] =&\frac{4}{c^4 P^N_4}\mathrm{E}[\mathrm{tr}(\mathbf X_i\mathbf X^{\prime}_j\mathbf X_k\mathbf X^{\prime}_l)\mathrm{tr}(\mathbf X_i\mathbf X^{\prime}_j\mathbf X_k\mathbf X^{\prime}_l)]\\
&+\frac{4}{c^4 P^N_4}\mathrm{E}[\mathrm{tr}(\mathbf X_i\mathbf X^{\prime}_j\mathbf X_k\mathbf X^{\prime}_l)\mathrm{tr}(\mathbf X_i\mathbf X^{\prime}_j\mathbf X_l\mathbf X^{\prime}_k)]\\
            &+\frac{4}{c^4 P^N_4}\mathrm{E}[\mathrm{tr}(\mathbf X_i\mathbf X^{\prime}_j\mathbf X_k\mathbf X^{\prime}_l)\mathrm{tr}(\mathbf X_i\mathbf X^{\prime}_k\mathbf X_j\mathbf X^{\prime}_l)]\\
            &+\frac{4}{c^4 P^N_4}\mathrm{E}[\mathrm{tr}(\mathbf X_i\mathbf X^{\prime}_j\mathbf X_k\mathbf X^{\prime}_l)\mathrm{tr}(\mathbf X_i\mathbf X^{\prime}_k\mathbf X_l\mathbf X^{\prime}_j)]\\
            &+\frac{4}{c^4 P^N_4}\mathrm{E}[\mathrm{tr}(\mathbf X_i\mathbf X^{\prime}_j\mathbf X_k\mathbf X^{\prime}_l)\mathrm{tr}(\mathbf X_i\mathbf X^{\prime}_l\mathbf X_j\mathbf X^{\prime}_k)]\\
            &+\frac{4}{c^4 P^N_4}\mathrm{E}[\mathrm{tr}(\mathbf X_i\mathbf X^{\prime}_j\mathbf X_k\mathbf X^{\prime}_l)\mathrm{tr}(\mathbf X_i\mathbf X^{\prime}_l\mathbf X_k\mathbf X^{\prime}_j)]\\
            &=\frac{4}{P^N_4} \left\{\frac{\mathrm{tr}^2(\boldsymbol \Sigma^2_C)}{c^4}\left[\mathrm{tr}^2(\boldsymbol \Sigma^2_R)+\mathrm{tr}(\boldsymbol \Sigma^4_R)\right]+\frac{\mathrm{tr}(\boldsymbol \Sigma^4_C)}{c^4}\left[\mathrm{tr}^2(\boldsymbol \Sigma^2_R)+3\mathrm{tr}(\boldsymbol \Sigma^4_R)\right]\right\}.       
\end{align*}
Finally, $\mathrm{E}[Y_{1N}Y_{3N}]=\mathrm{E}[Y_{1N}Y_{4N}]=\mathrm{E}[Y_{1N}Y_{5N}]=\mathrm{E}[Y_{2N}Y_{3N}]=\mathrm{E}[Y_{2N}Y_{4N}]=\mathrm{E}[Y_{2N}Y_{5N}]=\mathrm{E}[Y_{3N}Y_{4N}]=\mathrm{E}[Y_{3N}Y_{5N}]=\mathrm{E}[Y_{4N}Y_{5N}]=0$
and 
\begin{align*}
\mathrm{E}[Y_{1N}Y_{2N}]=&\frac{2}{c^{3}N}\mathrm{E}[\mathrm{tr}(\mathbf X_i \mathbf X^{\prime}_i)\mathrm{tr}(\mathbf X_i \mathbf X^{\prime}_i\mathbf X_j \mathbf X^{\prime}_j)]\\
                         &+\frac{N-2}{c^3N}\mathrm{E}[\mathrm{tr}(\mathbf X_i \mathbf X^{\prime}_i)\mathrm{tr}(\mathbf X_j \mathbf X^{\prime}_j\mathbf X_k \mathbf X^{\prime}_k)]\\
               =&\mathrm{tr}(\boldsymbol \Sigma_R^2) \mathrm{tr}(\boldsymbol \Sigma_R)+\frac{4}{N}\frac{\mathrm{tr}(\boldsymbol \Sigma_C^2)}{c^2}\mathrm{tr}(\boldsymbol \Sigma_R^3)\\
               &+\frac{2B}{N}\frac{\mathrm{tr}(\boldsymbol \Sigma_C \circ \boldsymbol \Sigma_C)}{c^2} \mathrm{tr}(\boldsymbol \Sigma^2_R \circ \boldsymbol \Sigma_R).
\end{align*}

\section{Proofs}
\begin{proof}
The essential step is to show that under model~(2.1) and assumption~(3.3)
$$\frac{G_N-\mathrm{E}[G_N]}{\mathrm{Var}[G_N]} \stackrel{d}{\rightarrow}\mathrm{N}(0,1),$$ 
where $G_N=\kappa_{1N}Y_{1N}+\kappa_{2N} Y_{2N}$ and $\kappa_{1N}$,$\kappa_{2N}$ are arbitrary constants. To accomplish this, the martingale central limit theorem will be used. Let $\mathcal{F}_0=\{\emptyset,\Omega\}$, $\mathcal{F}_k=\sigma\{\mathbf X_1,\ldots,\mathbf X_k\}$ for $k=1,\ldots,N$, $E_k$ be the conditional expectation given $\mathcal{F}_k$, $D_{Nk}=(E_k-E_{k-1})G_N$ and $S_{Nm}=\sum_{k=1}^m D_{Nk}=E_m[G_N]-\mathrm{E}[G_N]$. Write 
\begin{align*}
D_{Nk}&=\kappa_{1N}(E_k-E_{k-1})Y_{1N}+\kappa_{2N}(E_k-E_{k-1})Y_{2N}\\
      &=\frac{1}{cN}\left[\mathbf W^{\prime}_{k} \left(\boldsymbol \Sigma_C \otimes \boldsymbol \Lambda_N \right) \mathbf W_{k}-\mathrm{tr}\left(\boldsymbol \Sigma_C \otimes \boldsymbol \Lambda_N \right)\right]\\
      &+\frac{2\kappa_{2N}}{c^2 P^{N}_2}\left[\mathbf W^{\prime}_{k} \left(\boldsymbol \Sigma_C \otimes \mathbf M_{k-1} \right) \mathbf W_{k}-\mathrm{tr}\left(\boldsymbol \Sigma_C \otimes \mathbf M_{k-1} \right)\right]
\end{align*}
where $\mathbf W_{i}=\mathrm{vec}(\mathbf Z_{i})$, $\boldsymbol \Lambda_N=\kappa_{1N} \boldsymbol \Sigma_R+2\kappa_{2N} \boldsymbol \Sigma^2_R$, $\mathbf Q_{k}=\sum_{i=1}^{k} (\mathbf X_{i}\mathbf X^{\prime}_{i}-c \boldsymbol \Sigma_R)$ and $\mathbf M_{k}=\boldsymbol \Sigma^{1/2}_R \mathbf Q_{k} \boldsymbol \Sigma^{1/2}_R$. We need the following three lemmata:
\begin{lemma}
For any $N$, $\{D_{Nk},1\leq k \leq N\}$ is a martingale difference sequence with respect to the $\sigma$-fields $\{\mathcal{F}_k,1\leq k \leq N\}$.
\end{lemma}
\begin{proof}
Note that $\mathrm{E}[D_{Nk}]=0$ and write $S_{Nq}=S_{Nm}+E_q[G_N]-E_m[G_N]$ for $q>m$. Then it can be shown that $\mathrm{E}[S_{Nq}|\mathcal{F}_m]=S_{Nm}$ as desired.
\end{proof}
\begin{lemma}
Let $\sigma^2_{Nk}=E_{k-1}[D^2_{Nk}]$. Under assumption~(3.3) 
$$\frac{\sum_{k=1}^N \sigma^2_{Nk}}{\mathrm{Var}[G_N]} \stackrel{P}{\rightarrow} 1.$$
\end{lemma}
\begin{proof}
First note that 
\begin{align*}
\mathrm{Var}[G_N]=&\kappa^2_{1N} \mathrm{Var}[Y_{1N}]+\kappa^2_{2N} \mathrm{Var}[Y_{2N}]+2\kappa_{1N}\kappa_{2N}\mathrm{cov}[Y_{1N},Y_{2N}]\\        
        =&\frac{2}{N}\frac{\mathrm{tr}(\boldsymbol \Sigma^2_C)}{c^2}\mathrm{tr}(\boldsymbol \Lambda^2_N)+\frac{B}{N}\frac{\mathrm{tr}(\boldsymbol \Sigma_C \circ \boldsymbol \Sigma_C)}{c^2}\mathrm{tr}(\boldsymbol \Lambda_N \circ \boldsymbol \Lambda_N)\\
        &+\frac{4\kappa^2_{2N}}{N^2}\frac{\mathrm{tr}^2(\boldsymbol \Sigma_C^2)}{c^4} \mathrm{tr}^2(\boldsymbol \Sigma_R^2)\left\{1+O(N^{-1})\right\}.
\end{align*}
Next note that for large $N$, there exists a constant $\lambda_1$ such that
$$\left(\mathrm{Var}[G_N]\right)^2\geq \lambda_1\max\left\{\frac{\kappa^2_{2N}}{N^3}\frac{\mathrm{tr}^3(\boldsymbol \Sigma_C^2)}{c^6}\mathrm{tr}(\boldsymbol \Lambda^2_N)\mathrm{tr}^2(\boldsymbol \Sigma_R^2),\frac{\kappa^4_{2N}}{N^4}\frac{\mathrm{tr}^4(\boldsymbol \Sigma_C^2)}{c^8}\mathrm{tr}^4(\boldsymbol \Sigma_R)\right\}.$$
Next note that 
\begin{align*}
\sum_{k=1}^N \sigma^2_{Nk}=&\frac{8\kappa_{2N}}{NP^N_2}\frac{\mathrm{tr}(\boldsymbol \Sigma^2_C)}{c^2}\frac{1}{c} \sum_{k=1}^N(\kappa_{1N}\tau_{2(k-1)}+2\kappa_{2N}\tau_{3(k-1)})\\
                         &+\frac{4\kappa_{2N}B}{NP^N_2}\frac{\mathrm{tr}(\boldsymbol \Sigma_C \circ \boldsymbol \Sigma_C)}{c^2}\frac{1}{c} \sum_{k=1}^N \mathrm{tr}(\mathbf M_{k-1} \circ \boldsymbol  \Lambda_N)\\
                         &+\frac{8c^2_{2N}}{(P^N_2)^2}\frac{\mathrm{tr}(\boldsymbol \Sigma^2_C)}{c^2}\frac{1}{c^2} \sum_{k=1}^N \mathrm{tr}(\mathbf M^2_{k-1})\\
                         &+\frac{4Bc^2_{2N}}{(P^N_2)^2}\frac{\mathrm{tr}(\boldsymbol \Sigma_C \circ \boldsymbol \Sigma_C)}{c^2}\frac{1}{c^2} \sum_{k=1}^N \mathrm{tr}(\mathbf M_{k-1} \circ \mathbf M_{k-1})+H\\
      &=H_{1N}+H_{2N}+H_{3N}+H_{4N}+H,
\end{align*}
where $H$ is a finite constant, $\tau_{2k}=\sum_{i=1}^{k} \mathrm{tr}(\mathbf Q_{k} \boldsymbol \Sigma^2_R)$ and $\tau_{3k}=\sum_{i=1}^{k} \mathrm{tr}(\mathbf Q_{k} \boldsymbol \Sigma^3_R)$. To complete the proof, we need to show that $\mathrm{Var}[H_{Nm}]=o\left\{(\mathrm{Var}[G_N])^2\right\}$ for $m=1,2,3,4$. Note that when $k \leq j$
\begin{equation*}
\mathrm{cov}[\kappa_{1N} \tau_{2k}+2 \kappa_{2N} \tau_{3k}, \kappa_{1N} \tau_{2j}+2 \kappa_{2N} \tau_{3j}] = \mathrm{Var}[\kappa_{1N} \tau_{2k}+2 \kappa_{2N} \tau_{3k}]
\end{equation*} 
and
\begin{align*}
\mathrm{Var}[\kappa_{1N} \tau_{2k}+2 \kappa_{2N} \tau_{3k}]=&k \mathrm{Var}[\kappa_{1N} \mathrm{tr}(\mathbf X_i \mathbf X^{\prime}_i \boldsymbol\Sigma^2_R)+ 2 \kappa_{2N} \mathrm{tr}(\mathbf X_i \mathbf X^{\prime}_i \boldsymbol\Sigma^3_R)] \\
                                                            =&k \mathrm{Var}[\mathbf W^{\prime}_i (\boldsymbol \Sigma_C \otimes \boldsymbol \Lambda_N \boldsymbol\Sigma^2_R) \mathbf W_i] \\
                                                            =&2k  \mathrm{tr}(\boldsymbol \Sigma^2_C) \mathrm{tr}\left[(\boldsymbol \Lambda_N \boldsymbol \Sigma^2_R)^2\right]\\
                                                            &+B \mathrm{tr}(\boldsymbol \Sigma_C \circ \boldsymbol \Sigma_C) \mathrm{tr}\left[(\boldsymbol \Lambda_N \boldsymbol \Sigma_R^2) \circ (\boldsymbol \Lambda_N \boldsymbol \Sigma_R^2)\right] \\
                                                      \leq &k (2+\max\{0,B\}) \mathrm{tr}(\boldsymbol \Sigma^2_C) \mathrm{tr}(\boldsymbol \Lambda_N^2) \mathrm{tr}(\boldsymbol \Sigma^4_R)                                                     
\end{align*}
Therefore there exists a constant $\lambda_2$ such that 
\begin{equation*}
\frac{\mathrm{Var}[H_{N1}]}{(\mathrm{Var}[G_N])^2} \leq \frac{\lambda_2 \frac{\kappa^2_{2N}}{N^3}\left(\frac{\mathrm{tr}(\boldsymbol \Sigma^2_C)}{c^2}\right)^3 \mathrm{tr}(\boldsymbol \Lambda_N^2)\mathrm{tr}(\boldsymbol \Sigma^4_R)}{\lambda_1 \frac{\kappa^2_{2N}}{N^3}\left(\frac{\mathrm{tr}(\boldsymbol \Sigma^2_C)}{c^2}\right)^3 \mathrm{tr}(\boldsymbol \Lambda_N^2)\mathrm{tr}^2(\boldsymbol{\Sigma}^2_R)}=\frac{\lambda_2}{\lambda_1}\frac{\mathrm{tr}(\boldsymbol \Sigma^4_R)}{\mathrm{tr}^2(\boldsymbol{\Sigma}^2_R)}\rightarrow 0
\end{equation*}
as desired. Similar operations can show that $\mathrm{Var}[H_{Nm}]=o(\mathrm{Var}^2[G_N])$ for $m=2,3,4$.
\end{proof}
\begin{lemma}
Under assumption~(3.3) 
$$\frac{\sum_{k=1}^N \mathrm{E}[D^4_{Nk}]}{(\mathrm{Var}[G_N])^2}\rightarrow 0.$$
\end{lemma}
\begin{proof}
By the Cauchy-Schwarz inequality there exist constants $\lambda_3$ and $\lambda_4$ such that
\begin{align*}
\mathrm{E}[D^4_{Nk}]& \leq \lambda_3\frac{1}{c^4N^3}\mathrm{E}[\mathbf W^{\prime}_{k} \left(\boldsymbol \Sigma_C \otimes \boldsymbol \Lambda_N \right) \mathbf W_{k}-\mathrm{tr}\left(\boldsymbol \Sigma_C \otimes \boldsymbol \Lambda_N \right)]^4\\
           &+\lambda_4\frac{ 2 \kappa_{2N}}{p^8_2 (P^{N}_2)^4}\sum_{k=1}^N \mathrm{E}[\mathbf W^{\prime}_{k} \left(\boldsymbol \Sigma_C \otimes \mathbf M_{k-1} \right) \mathbf W_{k}-\mathrm{tr}\left(\boldsymbol \Sigma_C \otimes \mathbf M_{k-1} \right)]^4\\
           & \leq \frac{\lambda^{\prime}_3}{N^3} \frac{\mathrm{tr}^2(\boldsymbol \Sigma_C^2)}{c^4} \mathrm{tr}^2(\boldsymbol \Lambda_N^2) +\frac{\lambda^{\prime}_4}{N^6} \frac{\mathrm{tr}^2(\boldsymbol \Sigma_C^2)}{c^4} \mathrm{tr}^2(\boldsymbol \Sigma_R^4)
\end{align*}
Hence 
$$\frac{\sum_{k=1}^N \mathrm{E}[D^4_{Nk}]}{(\mathrm{Var}[G_N])^2} \leq \frac{\lambda^{\prime}_1}{cN}+\frac{\lambda^{\prime}_2}{cN^2} \frac{\mathrm{tr}^2(\boldsymbol \Sigma_R^4)}{\mathrm{tr}^4(\boldsymbol \Sigma_R^2)} \rightarrow 0,$$
for some constants $\lambda^{\prime}_1$, $\lambda^{\prime}_2$, $\lambda^{\prime}_3$ and $\lambda^{\prime}_4$.
\end{proof}
Combining the three lemmata it follows that $(G_N-\mathrm{E}[G_N])/\mathrm{Var}[G_N]\stackrel{d}{\rightarrow}\mathrm{N}(0,1)$. Next write 
$$\frac{\mathrm{tr}^2(\boldsymbol \Sigma_R)}{\mathrm{tr}(\boldsymbol \Sigma^2_R)} \frac{U_N+1}{r}-1=\frac{\tilde{U}_N-\tilde{T}_{1N}^2}{\left(1+\tilde{T}_{1N}\right)^2}$$
where $$\tilde{U}_N=\frac{T_{2N}}{\mathrm{tr}(\boldsymbol \Sigma^2_R)}-2\frac{T_{1N}}{\mathrm{tr}(\boldsymbol \Sigma_R)}+1 \text{ and } \tilde{T}_{1N}=\frac{T_{1N}-\mathrm{tr}(\boldsymbol \Sigma_R)}{\mathrm{tr}(\boldsymbol \Sigma_R)}.$$
To complete the proof of this theorem, we need to show that $\tilde{T}_{1N} \stackrel{P}{\rightarrow} 0$, $\sigma^{-1}_U \tilde{T}_{1N} \stackrel{P}{\rightarrow} 0$ and $\sigma^{-1}_U \tilde{U}_N \stackrel{d}{\rightarrow}\mathrm{N}(0,1)$. These results are established in a similar fashion as in the proof of Theorem 1 in \cite{Chen2010a}. Since 
\begin{align*}
 \mathrm{Var}[\tilde{T}_{1N}]&=\frac{2}{N-1}\frac{\mathrm{tr}(\boldsymbol \Sigma^2_C)}{c^2}\frac{\mathrm{tr}(\boldsymbol \Sigma^2_R)}{\mathrm{tr}^2(\boldsymbol \Sigma_R)}+\frac{B}{N}\frac{\mathrm{tr}(\boldsymbol \Sigma_C \circ \boldsymbol \Sigma_C)}{c^2} \frac{\mathrm{tr}(\boldsymbol \Sigma_R \circ \boldsymbol \Sigma_R)}{{\mathrm{tr}^2(\boldsymbol \Sigma_R)}}\\
                 &\leq \left[\frac{2}{N-1}+\frac{\max\{0,B\}}{N}\right]\frac{\mathrm{tr}(\boldsymbol \Sigma^2_C)}{c^2} \frac{\mathrm{tr}(\boldsymbol \Sigma^2_R)}{\mathrm{tr}^2(\boldsymbol \Sigma_R)},
                 \end{align*}
and $\mathrm{E}[\tilde{T}_{1N}]=0$, it follows that $\tilde{T}_{1N} \stackrel{P}{\rightarrow} 0$ and $\sigma^{-1}_U \tilde{T}_{1N} \stackrel{P}{\rightarrow} 0$. Finally, note that $\mathrm{Var}\left[\tilde{U}_N\right]=\sigma^2_U\left\{1+o(1)\right\}$. It therefore follows that $\sigma^{-1}_U \tilde{U}_N \stackrel{d}{\rightarrow}\mathrm{N}(0,1)$ as desired.
\end{proof}

\begin{proof}[Proof of Theorem 2]
Derivations in \cite{Chen2010a} imply that $\mathrm{E}[T^{\ast}_{2N}]=\mathrm{tr}(\boldsymbol \Omega^2)$ and $\mathrm{Var}[T^{\ast}_{2N}]/\mathrm{tr}^2(\boldsymbol \Omega^2) \rightarrow 0$. Therefore, $T^{\ast}_{2N}$ is a ratio-consistent estimator of $\mathrm{tr}(\boldsymbol \Omega^2)$. Similarly, the moment derivations in Section \ref{Moment Derivations}, imply that $T_{2N}$ is ratio-consistent estimator of $\mathrm{tr}(\boldsymbol \Sigma_R^2)$. The last claim of the theorem follows from the continuity mapping theorem.
\end{proof}

\begin{proof}[Proof of Theorem 3]
The proof is similar to that of Theorem 2 in \cite{Chen2010a}. Write $rV_N=(Y_{2N}-2Y_{1N}+r)+2Y_{3N}-2Y_{4N}+Y_{5N}$ and note that $\mathrm{E}[rV_N]=\mathrm{tr}\left[(\boldsymbol \Sigma_R-\mathbf I_{r})^2\right]$, $\mathrm{Var}[rV_N]=\sigma^2_V\left\{1+o(1)\right\}$. Therefore $$\frac{r V_N-\mathrm{tr}\left[(\boldsymbol \Sigma_R -\mathbf I_{r})^2\right]}{\sigma_V} \stackrel{d}{\rightarrow}\mathrm{N}(0,1)$$ as desired.
\end{proof}

\section{Simulation Results}
\indent Table~\ref{tab:size} contains the empirical levels of the proposed sphericity test for the two distributional scenarios under a weak and a strong column-wise correlation pattern. The sphericity test was slightly liberal for small values of $N$, $r$ or $c$ but the difference between the empirical and the nominal level diminished as $N$, $r$ and $c$ all increased due to the asymptotic nature of the proposed test. Conditional on $(N,r,c)$ and $\boldsymbol \Sigma_C$, the empirical levels were comparable under both distributional scenarios due to the non-parametric nature of the test statistic. In the sampling schemes with small $N$ and/or $r$ the empirical level was closer to the nominal when $\rho=0.85$ rather than when $\rho=0.15$. Hence, the proposed test does not confound a weak row-wise correlation pattern with a strong column-wise pattern but some attention is required when both correlation patterns are weak and the sample size is small. 

\indent Table~\ref{Power1} displays the empirical powers of the proposed sphericity test under the compound symmetry form for $\boldsymbol \Sigma_R$ in sampling schemes with a strong column-wise dependence structure, and Table~\ref{Power2} contains the empirical powers under the tridiagonal form for $\boldsymbol \Sigma_R$. We do not report the results for the compound symmetry structure since the empirical powers were almost all equal to $1.0$. We observed the following trends. First, the empirical powers were affected by the strength of the column-wise dependence structure, with weak correlation patterns boosting the empirical powers. According to the power analysis, this should be attributed to value of $\mathrm{tr}(\boldsymbol \Sigma^4_R)/\mathrm{tr}^2(\boldsymbol \Sigma^2_R)$ which converges to $0$ faster for the smaller value of $\rho$ while keeping the other parameters fixed. Second, the empirical powers approached $1.0$ as one or more of the elements in the triplet $(N,r,c)$ increased, indicating the consistency of the proposed tests under the working assumption that allows us to handle the `small $N$ large$p$' situation. Finally, no significant difference was noticed in the empirical powers in any of the two distributional scenarios. 

\indent Due to lack of alternative testing procedures and since $\mathrm{tr}(\boldsymbol \Sigma^2_C)/c^2 \in [1/c,1]$, we considered three alternative statistics by setting the `nuisance' ratio $\mathrm{tr}(\boldsymbol \Sigma^2_C)/c^2$ equal to the two boundary values and to the true value of this ratio. If $\mathrm{tr}(\boldsymbol \Sigma^2_C)/c^2=1/c$, the difference between the empirical and the nominal level of the resulting test statistic was small when $\rho=0.15$ but larger when $\rho=0.85$. Since $\mathrm{tr}(\boldsymbol \Sigma^2_C)=c$ is satisfied only when $\boldsymbol \Sigma_C=\mathbf I_{c}$, this explains the poor performance of the test statistic in the presence of a strong column-wise dependence structure. By contrast, if $\mathrm{tr}(\boldsymbol \Sigma^2_C)/c^2$ is set equal to 1, the resulting test becomes very conservative, failing to reject the sphericity hypothesis in all cases. For these reasons, we did not try to evaluate the empirical power of these two tests. Finally when we replaced $\mathrm{tr}(\boldsymbol \Sigma^2_C)/c^2$ with its true value, we did not observe any substantial difference with the results based on $U^{\ast}_N$. Hence, accurate estimation of the `nuisance' parameter $\mathrm{tr}(\boldsymbol \Sigma^2_C)$ seems to be crucial if we want to preserve the nominal level and, most importantly, $\widehat{\mathrm{tr}(\boldsymbol \Sigma_C^2)}$ serves this purpose.

\begin{sidewaystable}
\caption{\label{tab:size} Empirical levels of the proposed sphericity test for $H_0:\boldsymbol \Sigma_R=\sigma^2 \mathbf I_{r}$ versus $H_1:\boldsymbol \Sigma_R\neq \sigma^2 \mathbf I_{r}$ at $5\%$ nominal significance level.}
\begin{tabular*}{\textwidth}{c @{\extracolsep{\fill}}rrrrrrrrrrrrrrr}
  \toprule
    &       & \multicolumn{6}{c}{$\rho=0.15$} &&  \multicolumn{6}{c}{$\rho=0.85$}\\
    &       & \multicolumn{6}{c}{$r$} && \multicolumn{6}{c}{$r$}\\ 
    \cline{3-8} \cline{10-15}
$N$ & $c$ & 8 & 16 & 32 & 64 & 128 & 256 && 8 & 16 & 32 & 64 & 128 & 256 \\ 
  \midrule
      &       & \multicolumn{13}{c}{Scenario 1}  \\
 20 & 10  & 0.086 & 0.077 & 0.079 & 0.062 & 0.078 & 0.075 && 0.047 & 0.056 & 0.059 & 0.065 & 0.062 & 0.062 \\ 
    & 50  & 0.069 & 0.084 & 0.063 & 0.059 & 0.063 & 0.069 && 0.056 & 0.075 & 0.062 & 0.054 & 0.058 & 0.055 \\ 
    & 100 & 0.081 & 0.066 & 0.066 & 0.060 & 0.057 & 0.063 && 0.073 & 0.063 & 0.049 & 0.056 & 0.058 & 0.062 \\ 
 40 & 10  & 0.069 & 0.060 & 0.059 & 0.069 & 0.057 & 0.059 && 0.063 & 0.061 & 0.055 & 0.058 & 0.060 & 0.058 \\ 
    & 50  & 0.059 & 0.056 & 0.070 & 0.050 & 0.057 & 0.072 && 0.045 & 0.055 & 0.059 & 0.066 & 0.055 & 0.053 \\ 
    & 100 & 0.067 & 0.046 & 0.061 & 0.051 & 0.045 & 0.054 && 0.060 & 0.056 & 0.061 & 0.053 & 0.058 & 0.065 \\ 
 60 & 10  & 0.067 & 0.060 & 0.068 & 0.074 & 0.051 & 0.054 && 0.068 & 0.056 & 0.062 & 0.067 & 0.076 & 0.055 \\ 
    & 50  & 0.058 & 0.075 & 0.056 & 0.060 & 0.049 & 0.058 && 0.059 & 0.064 & 0.055 & 0.058 & 0.049 & 0.052 \\ 
    & 100 & 0.066 & 0.050 & 0.050 & 0.047 & 0.043 & 0.045 && 0.061 & 0.055 & 0.058 & 0.062 & 0.058 & 0.064 \\ 
 80 & 10  & 0.081 & 0.057 & 0.058 & 0.070 & 0.050 & 0.047 && 0.072 & 0.050 & 0.055 & 0.057 & 0.049 & 0.058 \\ 
    & 50  & 0.060 & 0.059 & 0.051 & 0.046 & 0.047 & 0.058 && 0.057 & 0.053 & 0.045 & 0.058 & 0.056 & 0.000 \\ 
    & 100 & 0.065 & 0.064 & 0.046 & 0.050 & 0.050 & 0.055 && 0.048 & 0.052 & 0.071 & 0.045 & 0.053 & 0.045 \\ 
          &       & \multicolumn{13}{c}{Scenario 2}  \\
 20 & 10  & 0.097 & 0.087 & 0.079 & 0.069 & 0.082 & 0.066 && 0.064 & 0.068 & 0.051 & 0.059 & 0.061 & 0.055 \\ 
    & 50  & 0.088 & 0.079 & 0.057 & 0.059 & 0.058 & 0.067 && 0.081 & 0.068 & 0.052 & 0.067 & 0.054 & 0.063 \\ 
    & 100 & 0.085 & 0.072 & 0.062 & 0.066 & 0.070 & 0.066 && 0.072 & 0.067 & 0.048 & 0.070 & 0.067 & 0.079 \\ 
 40 & 10  & 0.086 & 0.079 & 0.063 & 0.065 & 0.056 & 0.052 && 0.063 & 0.063 & 0.052 & 0.050 & 0.050 & 0.052 \\ 
    & 50  & 0.073 & 0.073 & 0.062 & 0.060 & 0.048 & 0.050 && 0.071 & 0.057 & 0.052 & 0.047 & 0.062 & 0.053 \\ 
    & 100 & 0.067 & 0.067 & 0.063 & 0.056 & 0.046 & 0.058 && 0.068 & 0.067 & 0.054 & 0.056 & 0.055 & 0.077 \\ 
 60 & 10  & 0.096 & 0.076 & 0.055 & 0.055 & 0.062 & 0.057 && 0.055 & 0.051 & 0.045 & 0.047 & 0.049 & 0.053 \\ 
    & 50  & 0.082 & 0.070 & 0.066 & 0.056 & 0.049 & 0.055 && 0.067 & 0.059 & 0.047 & 0.060 & 0.054 & 0.070 \\ 
    & 100 & 0.070 & 0.078 & 0.075 & 0.077 & 0.048 & 0.062 && 0.058 & 0.064 & 0.061 & 0.058 & 0.063 & 0.054 \\ 
 80 & 10  & 0.074 & 0.074 & 0.049 & 0.055 & 0.044 & 0.056 && 0.052 & 0.072 & 0.055 & 0.046 & 0.043 & 0.058 \\ 
    & 50  & 0.083 & 0.064 & 0.051 & 0.053 & 0.057 & 0.067 && 0.070 & 0.058 & 0.045 & 0.061 & 0.064 & 0.057 \\ 
    & 100 & 0.070 & 0.054 & 0.049 & 0.047 & 0.046 & 0.057 && 0.056 & 0.045 & 0.056 & 0.051 & 0.052 & 0.052 \\  
    \bottomrule
\end{tabular*}
\end{sidewaystable}

\begin{sidewaystable}
\caption{Empirical powers of the proposed sphericity test for $H_0:\boldsymbol \Sigma_R=\sigma^2 \mathbf I_{r}$ versus $H_1:\boldsymbol \Sigma_R=diag(\mathbf 2_{[r/8]},\mathbf 1_{[7r/8]})$ at $5\%$ nominal significance level.}
\begin{tabular*}{\textwidth}{c @{\extracolsep{\fill}} rrrrrrrrrrrrrrr}
  \toprule
    &       & \multicolumn{6}{c}{$\rho=0.15$} &&  \multicolumn{6}{c}{$\rho=0.85$}\\
    &       & \multicolumn{6}{c}{$r$} && \multicolumn{6}{c}{$r$}\\ 
    \cline{3-8} \cline{10-15}
$N$ & $c$ & 8 & 16 & 32 & 64 & 128 & 256 && 8 & 16 & 32 & 64 & 128 & 256 \\ 
  \midrule
      &       & \multicolumn{13}{c}{Scenario 1}  \\
20 & 10  & 0.987 & 0.998 & 1.000 & 1.000 & 1.000 & 1.000 && 0.458 & 0.512 & 0.559 & 0.582 & 0.544 & 0.586 \\ 
   & 50  & 1.000 & 1.000 & 1.000 & 1.000 & 1.000 & 1.000 && 0.988 & 0.999 & 0.999 & 1.000 & 1.000 & 1.000 \\ 
   & 100 & 1.000 & 1.000 & 1.000 & 1.000 & 1.000 & 1.000 && 1.000 & 1.000 & 1.000 & 1.000 & 1.000 & 1.000 \\ 
40 & 10  & 1.000 & 1.000 & 1.000 & 1.000 & 1.000 & 1.000 && 0.814 & 0.863 & 0.935 & 0.945 & 0.959 & 0.982 \\ 
   & 50  & 1.000 & 1.000 & 1.000 & 1.000 & 1.000 & 1.000 && 1.000 & 1.000 & 1.000 & 1.000 & 1.000 & 1.000 \\ 
   & 100 & 1.000 & 1.000 & 1.000 & 1.000 & 1.000 & 1.000 && 1.000 & 1.000 & 1.000 & 1.000 & 1.000 & 1.000 \\ 
60 & 10  & 1.000 & 1.000 & 1.000 & 1.000 & 1.000 & 1.000 && 0.951 & 0.981 & 0.996 & 0.998 & 1.000 & 1.000 \\ 
   & 50  & 1.000 & 1.000 & 1.000 & 1.000 & 1.000 & 1.000 && 1.000 & 1.000 & 1.000 & 1.000 & 1.000 & 1.000 \\ 
   & 100 & 1.000 & 1.000 & 1.000 & 1.000 & 1.000 & 1.000 && 1.000 & 1.000 & 1.000 & 1.000 & 1.000 & 1.000 \\ 
80 & 10  & 1.000 & 1.000 & 1.000 & 1.000 & 1.000 & 1.000 && 0.988 & 0.999 & 1.000 & 1.000 & 1.000 & 1.000 \\ 
   & 50  & 1.000 & 1.000 & 1.000 & 1.000 & 1.000 & 1.000 && 1.000 & 1.000 & 1.000 & 1.000 & 1.000 & 1.000 \\ 
   & 100 & 1.000 & 1.000 & 1.000 & 1.000 & 1.000 & 1.000 && 1.000 & 1.000 & 1.000 & 1.000 & 1.000 & 1.000\\ 
         &       & \multicolumn{13}{c}{Scenario 2}  \\
20 & 10  & 0.958 & 0.990 & 1.000 & 1.000 & 1.000 & 1.000 && 0.435 & 0.496 & 0.546 & 0.530 & 0.586 & 0.584 \\ 
   & 50  & 1.000 & 1.000 & 1.000 & 1.000 & 1.000 & 1.000 && 0.978 & 0.995 & 1.000 & 1.000 & 1.000 & 1.000 \\ 
   & 100 & 1.000 & 1.000 & 1.000 & 1.000 & 1.000 & 1.000 && 1.000 & 1.000 & 1.000 & 1.000 & 1.000 & 1.000 \\ 
40 & 10  & 1.000 & 1.000 & 1.000 & 1.000 & 1.000 & 1.000 && 0.784 & 0.870 & 0.920 & 0.945 & 0.962 & 0.980 \\ 
   & 50  & 1.000 & 1.000 & 1.000 & 1.000 & 1.000 & 1.000 && 1.000 & 1.000 & 1.000 & 1.000 & 1.000 & 1.000 \\ 
   & 100 & 1.000 & 1.000 & 1.000 & 1.000 & 1.000 & 1.000 && 1.000 & 1.000 & 1.000 & 1.000 & 1.000 & 1.000 \\ 
60 & 10  & 1.000 & 1.000 & 1.000 & 1.000 & 1.000 & 1.000 && 0.928 & 0.979 & 0.993 & 0.997 & 0.999 & 0.999 \\ 
   & 50  & 1.000 & 1.000 & 1.000 & 1.000 & 1.000 & 1.000 && 1.000 & 1.000 & 1.000 & 1.000 & 1.000 & 1.000 \\ 
   & 100 & 1.000 & 1.000 & 1.000 & 1.000 & 1.000 & 1.000 && 1.000 & 1.000 & 1.000 & 1.000 & 1.000 & 1.000 \\ 
80 & 10  & 1.000 & 1.000 & 1.000 & 1.000 & 1.000 & 1.000 && 0.986 & 0.991 & 1.000 & 1.000 & 1.000 & 1.000 \\ 
   & 50  & 1.000 & 1.000 & 1.000 & 1.000 & 1.000 & 1.000 && 1.000 & 1.000 & 1.000 & 1.000 & 1.000 & 1.000 \\ 
   & 100 & 1.000 & 1.000 & 1.000 & 1.000 & 1.000 & 1.000 && 1.000 & 1.000 & 1.000 & 1.000 & 1.000 & 1.000 \\ 
        \bottomrule
\end{tabular*}
\label{Power1}
\end{sidewaystable}

\begin{sidewaystable}
\caption{Empirical powers of the proposed sphericity test for $H_0:\boldsymbol \Sigma_R=\sigma^2 \mathbf I_{r}$ versus $H_1:\boldsymbol \Sigma_R=\{0.1^{|a-b|}I(|a-b|\leq 1)\}_{1\leq a,b\leq r}$ at $5\%$ nominal significance level.}
\begin{tabular*}{\textwidth}{c @{\extracolsep{\fill}} rrrrrrrrrrrrrrr}
  \toprule
    &       & \multicolumn{6}{c}{$\rho=0.15$} &&  \multicolumn{6}{c}{$\rho=0.85$}\\
    &       & \multicolumn{6}{c}{$r$} && \multicolumn{6}{c}{$r$}\\ 
    \cline{3-8} \cline{10-15}
$N$ & $c$ & 8 & 16 & 32 & 64 & 128 & 256 && 8 & 16 & 32 & 64 & 128 & 256 \\ 
 \midrule
      &       & \multicolumn{13}{c}{Scenario 1}  \\
20 & 10  & 0.448 & 0.499 & 0.565 & 0.580 & 0.595 & 0.612 && 0.112 & 0.123 & 0.115 & 0.130 & 0.136 & 0.130 \\ 
   & 50  & 1.000 & 1.000 & 1.000 & 1.000 & 1.000 & 1.000 && 0.383 & 0.471 & 0.493 & 0.481 & 0.492 & 0.500 \\ 
   & 100 & 1.000 & 1.000 & 1.000 & 1.000 & 1.000 & 1.000 && 1.000 & 1.000 & 1.000 & 1.000 & 1.000 & 1.000 \\ 
40 & 10  & 0.804 & 0.909 & 0.948 & 0.974 & 0.967 & 0.981 && 0.200 & 0.230 & 0.223 & 0.225 & 0.221 & 0.247 \\ 
   & 50  & 1.000 & 1.000 & 1.000 & 1.000 & 1.000 & 1.000 && 0.771 & 0.859 & 0.887 & 0.932 & 0.940 & 0.943 \\ 
   & 100 & 1.000 & 1.000 & 1.000 & 1.000 & 1.000 & 1.000 && 1.000 & 1.000 & 1.000 & 1.000 & 1.000 & 1.000 \\ 
60 & 10  & 0.971 & 0.995 & 0.997 & 1.000 & 1.000 & 1.000 && 0.308 & 0.341 & 0.366 & 0.390 & 0.388 & 0.359 \\ 
   & 50  & 1.000 & 1.000 & 1.000 & 1.000 & 1.000 & 1.000 && 0.951 & 0.986 & 0.994 & 0.997 & 1.000 & 1.000 \\ 
   & 100 & 1.000 & 1.000 & 1.000 & 1.000 & 1.000 & 1.000 && 1.000 & 1.000 & 1.000 & 1.000 & 1.000 & 1.000 \\ 
80 & 10  & 0.996 & 1.000 & 1.000 & 1.000 & 1.000 & 1.000 && 0.409 & 0.480 & 0.517 & 0.538 & 0.570 & 0.545 \\ 
   & 50  & 1.000 & 1.000 & 1.000 & 1.000 & 1.000 & 1.000 && 0.988 & 0.999 & 1.000 & 1.000 & 1.000 & 1.000 \\ 
   & 100 & 1.000 & 1.000 & 1.000 & 1.000 & 1.000 & 1.000 && 1.000 & 1.000 & 1.000 & 1.000 & 1.000 & 1.000 \\ 
    &       & \multicolumn{13}{c}{Scenario 2}  \\
20 & 10  & 0.449 & 0.522 & 0.527 & 0.567 & 0.593 & 0.579 && 0.114 & 0.141 & 0.127 & 0.120 & 0.146 & 0.117 \\ 
   & 50  & 1.000 & 1.000 & 1.000 & 1.000 & 1.000 & 1.000 && 0.387 & 0.449 & 0.447 & 0.499 & 0.502 & 0.523 \\ 
   & 100 & 1.000 & 1.000 & 1.000 & 1.000 & 1.000 & 1.000 && 1.000 & 1.000 & 1.000 & 1.000 & 1.000 & 1.000 \\ 
40 & 10  & 0.805 & 0.886 & 0.942 & 0.965 & 0.968 & 0.981 && 0.213 & 0.221 & 0.225 & 0.208 & 0.218 & 0.215 \\ 
   & 50  & 1.000 & 1.000 & 1.000 & 1.000 & 1.000 & 1.000 && 0.767 & 0.843 & 0.905 & 0.940 & 0.950 & 0.939 \\ 
   & 100 & 1.000 & 1.000 & 1.000 & 1.000 & 1.000 & 0.000 && 1.000 & 1.000 & 1.000 & 1.000 & 1.000 & 1.000 \\ 
60 & 10  & 0.950 & 0.990 & 0.998 & 1.000 & 0.999 & 1.000 && 0.311 & 0.345 & 0.365 & 0.379 & 0.360 & 0.384 \\ 
   & 50  & 1.000 & 1.000 & 1.000 & 1.000 & 1.000 & 1.000 && 0.938 & 0.989 & 0.998 & 1.000 & 0.999 & 1.000 \\ 
   & 100 & 1.000 & 1.000 & 1.000 & 1.000 & 1.000 & 0.000 && 1.000 & 1.000 & 1.000 & 1.000 & 1.000 & 1.000 \\ 
80 & 10  & 0.992 & 0.999 & 1.000 & 1.000 & 1.000 & 1.000 && 0.421 & 0.474 & 0.503 & 0.518 & 0.529 & 0.562 \\ 
   & 50  & 1.000 & 1.000 & 1.000 & 1.000 & 1.000 & 1.000 && 0.983 & 0.999 & 1.000 & 1.000 & 1.000 & 1.000 \\ 
   & 100 & 1.000 & 1.000 & 1.000 & 1.000 & 1.000 & 0.000 && 1.000 & 1.000 & 1.000 & 1.000 & 1.000 & 1.000 \\ 
    \bottomrule
\end{tabular*}
\label{Power2}
\end{sidewaystable}

\bibliographystyle{plainnat}
\bibliography{dissrefer.bib}